\documentclass[sigconf,authorversion]{acmart}

\usepackage{acmart-taps}
\usepackage{multirow}
\usepackage{xcolor,colortbl}
\usepackage{makecell}

\copyrightyear{2026}
\acmYear{2026}
\setcopyright{cc}
\setcctype{by}
\acmConference[CHI '26]{Proceedings of the 2026 CHI Conference on Human Factors in Computing Systems}{April 13--17, 2026}{Barcelona, Spain}
\acmBooktitle{Proceedings of the 2026 CHI Conference on Human Factors in Computing Systems (CHI '26), April 13--17, 2026, Barcelona, Spain}
\acmDOI{10.1145/3772318.3790520}
\acmISBN{979-8-4007-2278-3/2026/04}

\acmSubmissionID{6064}

\definecolor{colorR}{RGB}{255, 204, 204} 
\definecolor{colorA}{RGB}{255, 255, 204} 
\definecolor{colorC}{RGB}{204, 230, 255} 
\definecolor{colorI}{RGB}{220, 220, 220} 

\definecolor{phaseDesign}{RGB}{239, 224, 255} 
\definecolor{phaseEarlyDev}{RGB}{204, 255, 229} 
\definecolor{phaseMidDev}{RGB}{204, 242, 242} 
\definecolor{phasePostLaunch}{RGB}{255, 229, 204} 

\definecolor{darkcyan}{rgb}{0.07,0.51,0.77}
\definecolor{gold}{RGB}{255,194,0}
\definecolor{darkgreen}{RGB}{0,100,0}

\newcommand{\changed}[1]{{#1}}

\begin{document}

\title[Evaluating 3D A11y Guidelines with Industry XR Practitioners]{How Well Can 3D Accessibility Guidelines Support XR Development? An Interview Study with XR Practitioners in Industry}

\settopmatter{authorsperrow=4}

\author{Daniel Killough}
\email{dkillough@wisc.edu}
\orcid{0009-0002-2623-0528}
\affiliation{%
  \institution{University of Wisconsin-Madison}
  \country{United States}
}

\author{Tiger F. Ji}
\email{reatreify@gmail.com}
\affiliation{%
  \institution{University of Wisconsin-Madison}
  \country{United States}
}

\author{Kexin Zhang}
\email{kzhang284@wisc.edu}
\orcid{0009-0009-4078-8780}
\affiliation{%
  \institution{University of Wisconsin-Madison}
  \country{United States}
}

\author{Yaxin Hu}
\email{yaxin.hu@wisc.edu}
\affiliation{%
  \institution{University of Wisconsin-Madison}
  \country{United States}
}

\author{Yu Huang}
\email{yu.huang@vanderbilt.edu}
\affiliation{%
  \institution{Vanderbilt University}
  \city{Nashville}
  \state{Tennessee}
  \country{United States}
}

\author{Ruofei Du}
\email{me@duruofei.com}
\affiliation{%
  \institution{Google XR Labs}
  \city{San Francisco}
  \state{California}
  \country{United States}
}

\author{Yuhang Zhao}
\email{yuhang.zhao@cs.wisc.edu}
\affiliation{%
  \institution{University of Wisconsin-Madison}
  \country{United States}
}

\renewcommand{\shortauthors}{Killough, et al.}

\begin{abstract}
  
While accessibility (a11y) guidelines exist for 3D games and virtual worlds, their applicability to extended reality (XR)'s unique interaction paradigms (\textit{e.g.,} spatial tracking, kinesthetic interactions) remains unexplored. XR practitioners need practical guidance to successfully implement a11y guidelines under real-world constraints. We present the first evaluation of existing 3D a11y guidelines applied to XR development through semi-structured interviews with 25 XR practitioners across diverse organization contexts.
We assessed 20 commonly-agreed a11y guidelines from six major resources  across visual, motor, cognitive, speech, and hearing domains, comparing practitioners' development practices against guideline applicability to XR. Our investigation reveals that guidelines can be highly effective when designed as transformation catalysts rather than compliance checklists, but fundamental mismatches exist between existing 3D guidelines and XR requirements, creating both implementation barriers and design gaps. This work provides foundational insights towards developing a11y guidelines and support tools that address XR's distinct characteristics. 

\end{abstract}



\maketitle

\section{Introduction}


Extended reality (XR) technologies create immersive experiences that fundamentally differ from traditional 2D interfaces \changed{or 3D virtual worlds} displayed on flat screens. XR's unique interaction paradigms, including kinesthetic input, novel spatial interactions, and embodied sense of presence, introduce unique accessibility (a11y) challenges that existing guidelines fail to address and fundamentally alter how a11y must be conceptualized and implemented~\cite{heilemann2021guidelines, wang2025understanding}. 

While recently organizations have started developing a11y guidelines for 3D virtual worlds (\textit{e.g.,} \textit{Virtual Environments Accessibility Guidelines}~\cite{usc_ve_accessibility}, \textit{W3CXR}~\cite{w3cxr}, 
\textit{Game Accessibility Guidelines}~\cite{gameguide}), these frameworks were developed primarily from virtual experiences on 2D screens rather than the diverse XR ecosystem spanning virtual reality (VR), augmented reality (AR), and mixed reality (MR). More critically, no published work has systematically explored how XR practitioners---the main force who put guidelines into practice---experience, interpret, and adapt these a11y guidelines to XR development realities. This evaluation gap hinders widespread a11y adoption, as guidelines that practitioners cannot understand, implement, or see as relevant to their work will fail to improve a11y outcomes regardless of their theoretical soundness~\cite{colusso2017gap}. 

While recent work by Wang et al.~\cite{wang2025understanding} has identified organizational and motivational barriers to XR a11y implementation, practitioners actively seeking to create inclusive experiences lack adequate support methods and tools~\cite{colusso2017gap, zhang2025avatarguidelines}.
The gap between practitioner intent and available resources motivates our investigation into what enables XR a11y implementation. Understanding what support methods (\textit{e.g.,} guidelines, toolkits) work and why others fail is critical for enabling practitioners to implement a11y within real-world development constraints. 

We evaluated 20 a11y guidelines from established virtual world a11y resources with 25 XR practitioners, uncovering specific technical incompatibilities and implementation barriers unique to XR. 
Our investigation employs a semi-structured interview approach where guidelines serve as evaluation ``techniques'' that practitioners apply to their own development practices. This methodology creates a dual evaluation approach where practitioners simultaneously assess their development practices against guidelines while reflecting on the guidelines themselves for clarity, actionability, and applicability. \changed{Beyond guideline evaluation, we explored practitioners' perspectives on additional a11y support tools (\textit{e.g.,} toolkits, 3rd party plugins, automated a11y checking) to understand what support methods best enable efficient a11y implementation within typical XR development constraints.} Through this lens we explore the human experiences behind guideline interpretation and implementation, uncovering how professional communities construct meaning around a11y requirements and navigate competing values within XR development cultures. 

Our research addresses three questions through qualitative investigation: 
\begin{itemize}
    \item[\textbf{RQ1.}] What technical solutions are XR practitioners currently employing for a11y? What technical barriers prevent direct guideline implementation in XR?
    \item[\textbf{RQ2.}] How do XR practitioners interpret existing 3D a11y guidelines when applied to immersive contexts? 
    \item[\textbf{RQ3.}] What adaptations would make guidelines actionable for XR development? What additional support methods would enable efficient a11y integration within development constraints?
\end{itemize}

Our findings revealed: (1) fundamental tensions between immersive design values and a11y implementation that create unique professional challenges; (2) community practices for interpreting and adapting 3D guidelines to XR contexts; (3) technical and methodological incompatibilities between existing 3D guidelines and XR interaction paradigms; and (4) practitioners' recommendations for a11y support methods to integrate with existing XR development workflows (\textit{e.g,} improvements for more actionable guidelines). 

\changed{This work contributes the first 
exploration of how XR practitioners experience a11y guidelines in their professional contexts,} providing empirical evidence for improving guideline design, developing better a11y support tools, and understanding the social and cultural factors that influence accessible XR development.



\section{Related Work}


\label{relatedwork}
Our work examines how XR practitioners evaluate existing a11y guidelines, revealing systematic gaps in their applicability to immersive contexts and the social factors influencing implementation decisions. We contextualize our work within existing a11y guidelines for \changed{3D virtual worlds} and XR as well as prior literature on guideline implementation in traditional development contexts. 

\subsection{A11y Guidelines for Virtual Worlds and XR}
People with disabilities (PWD) face diverse challenges in XR, including kinesthetic interactions for wheelchair users~\cite{gerling2020virtual}, unpredictability of VR content increasing concerns for people with photosensitivity~\cite{south2024barriers}, and navigating social environments~\cite{collins2024ai}. Despite extensive assistive technology development for XR (\textit{e.g.,}~\cite{choi2018claw,heuten2006city,mott2020motor,nair2021navstick,picinali2014exploration,yamagami2021two,zhao2019seeingvr}), most research prototypes remain disconnected from industry practice, with
only a small number of 3D games having incorporated a11y features beyond basic settings like colorblind options (\textit{e.g.,}~\cite{lastofus,wbgames2024accessibility,worldsedge2024accessibility}), highlighting the barriers of large-scale a11y integration in industry. 

To support a11y integration in mainstream technologies, researchers and industry developers have created a11y guidelines to serve as foundational resources directing practitioners in implementing inclusive design features. Standard a11y guidelines have been constructed for web~\cite{brajnik2009valid, kurniawan2005older, mozilla, w3web} and mobile devices~\cite{w3mobile}, and have been used in various studies evaluating mobile platforms~\cite{ballantyne2018mobile, androidaccess, wei2018heuristic}. Dedicated a11y guidelines for 3D virtual worlds have also been constructed~\cite{apx, gameguide, xbox} alongside research efforts for more specific use cases (\textit{e.g.,} movement-based VR games~\cite{mason2022including, mueller2014movement} or expressing cultural heritage~\cite{a11ymuseum, chong2021virtual}). 
For example, the \textit{Game Accessibility Guidelines} (GAG)~\cite{gameguide} organize recommendations by disability type (motor, visual, cognitive, speech \& hearing) and provide implementation examples ranging from high-level design principles to concrete technical implementations. Similarly, the \textit{Xbox Accessibility Guidelines} (Xbox)~\cite{xbox} and \textit{Accessible Player Experiences} (APX)~\cite{apx} offer comprehensive frameworks for making 3D games more inclusive. These guidelines emerged primarily from desktop-based virtual worlds where 3D environments are viewed on flat screens with traditional input devices.
These 3D a11y guidelines have potential to be applied to XR accessibility as they address challenges absent from solely 2D interface standards (\textit{e.g.,} spatial audio design; depth perception requirements; motion-based interactions; 3D information presentation). However, they may also face fundamental limitations in XR contexts due to XR's unique interaction paradigms, such as embodied first-person experience and spatial, kinesthetic interactions~\cite{gerling2020virtual}. It is therefore unclear whether and how the existing 3D a11y guidelines may transfer to the emerging XR applications, for example, which recommendations may transfer effectively and which require fundamental reconceptualization for XR's unique interaction paradigms.

\changed{Beyond 3D a11y guidelines,} researchers~\cite{heilemann2021guidelines, berkeleyAccessibilityUniversal, mott2019design, melbourne} and organizations~\cite{xraA11y, usabilityheur, magicleap, oculusvr} have started compiling a11y guidelines specifically for XR experiences. Heilemann et al.~\cite{heilemann2021guidelines} compiled VR-relevant guidelines from existing game a11y resources, while organizations like Meta~\cite{oculusvr} and Magic Leap~\cite{magicleap} created device-specific recommendations for their platforms. Despite these ongoing efforts, current XR a11y guidelines are in their infancy, without rigorous validation or broad agreement. Moreover, these frameworks \textbf{mostly focus on accommodating an end-user rather than technical guidance to developers in how to implement these guidelines; they also remain largely untested with the practitioners responsible for implementing them, creating uncertainty about their practical applicability within real-world XR development contexts.} To our knowledge, no prior research has systematically validated the applicability and issues of existing 3D and XR a11y guidelines, and examined whether XR practitioners can successfully interpret and implement these guidelines when developing for immersive platforms.

\subsection{Implementing Guidelines}
Researchers have identified challenges of incorporating a11y into traditional development \cite{bai2017cost, bi2021github, bi2022pract, miranda2022agile}, revealing substantial variance in a11y development quality across developers and highlighting how professional identity and organizational context influence implementation decisions~\cite{bi2022pract} and adding PWD-focused user testing~\cite{bai2017cost}. 
Understanding how practitioners experience a11y guidelines has emerged as critical for effective implementation; established guideline evaluation methodologies demonstrate that practitioners can effectively assess both their practices against guidelines and evaluate guideline quality when provided with structured frameworks~\cite{ballantyne2018mobile,brajnik2009valid,wei2018heuristic}. 
These approaches have successfully validated web and mobile a11y standards while revealing critical insights about guideline design and implementation patterns. 

For XR specifically, Wang et al.'s investigation~\cite{wang2025understanding} represents the most comprehensive examination of VR practitioners' general a11y perspectives, revealing challenges including hardware limitations, insufficient professional knowledge, and competing development priorities. Additional research has explored a11y challenges in XR environments~\cite{creed2024inclusive,naikar2024accessibility} and broader XR development challenges~\cite{ashtari2020xr,krau2021collab}. While these foundational works establish understanding of practitioner attitudes toward a11y, they focus on general perspectives rather than evaluating specific guidelines.
\textbf{Our work extends this foundation by focusing specifically on how practitioners encounter, interpret, and evaluate existing a11y guidelines rather than exploring general attitudes and barriers toward a11y.} Beyond interviews about a11y challenges, we asked XR practitioners to systematically examine existing XR-relevant a11y guidelines and discuss their interpretation, potential use, and implementation concerns through shared projects in their preferred work environment.



\section{Method}


\label{methods}

We compiled existing 3D a11y guidelines and conducted 25 semi-structured interviews with XR practitioners to evaluate guideline feasibility and applicability to XR development contexts. 
Practitioners assessed both their development practices against guidelines and evaluated the guidelines themselves for XR applicability, creating a dual evaluation that reveals both technical implementation barriers and guideline design gaps specific to XR contexts.

\subsection{Participants}


\begin{table*}[ht]
\centering
\small
\Description{A table showing abridged demographic information of the participants. From left to right, column titles are PID, Gender, Company Size, Years of XR Development Experience and Primary Roles, Primary Dev Platforms, Prior A11y Experience, and Eval. In the first and second columns, participants are numbered 1 through 25, with P2 and P10 listed as Female and the rest Male. For company size, participants have overlapping experience, but the following participants are listed as big tech: P3, P5, P15, P19, P21, P23, P24, and P25; the following participants are listed as midsize: P7, P9, P10, P14, P16, P17, and P24. The following participants are listed as startup: P1, P2, P4, P6, P8, P11, P12, P13, P16, P18, P20, P21, P22, P23. The following participants are listed as freelance: P1, P2, P4, P5, P6, P8, P11, P14, P22, P23. In the fourth column, XR Development Experience and Primary Roles, participants are listed from less than 1 year for P17, on an a11y team at engine, then 2 years for P9, and up to 9 years for P8. The most participants have 4 to 6 years of experience. All participants except P12 and P18 have developer listed, who have designer listed instead. These participants only have developer listed: P2, P5, P9, P10, P11, P13, P15, P16, P17, P19, P21, and P24. Other participants have designer, researcher, or some sort of management role listed in addition to developer. In the fifth column, Primary Dev Platform(s), Unity is overwhelmingly listed, with few other options, being: BabylonJS Playground for P3; Glitch for P7 and P9; Custom Engine AND Unity for P8; Unreal Engine for P11; PlayCanvas for P14; P17's place of employment (redacted); and both Custom Engine AND Unreal for P25. All other rows just list Unity. In the sixth column, Prior A11y Exp, lists either ``no" or "yes''. If yes, the type of a11y impairment is listed or simply ``XR tools''.  With some overlapping experience, 20 participants are listed as 'yes', with 10 participants listed with visual experience: P2, P3, P4, P7, P13, P15, P19, P20, P23, and P24. 7 participants listed with motor experience: P1, P2, P4, P13, P19, P21, and P23. 5 participants listed with cognitive experience: P1, P7, P16, P22, and P25. 4 participants listed with speech and hearing experience: P5, P12, P23, and P25. 4 participants listed with XR tools experience: P3, P6, P15, and P16. 5 participants are listed as no a11y development experience: P8, P9, P10, P11, and P18. The final column, Eval, indicates the guideline category each participant evaluated: Visual, Cognitive, Motor, or Speech and Hearing (S and H).}
\resizebox{\textwidth}{!}{\begin{tabular}{@{}lcllll l@{}}
\toprule
\textbf{PID} & \textbf{Gender} & \textbf{Company Size Exp} & \textbf{Years XR Exp: Primary Role(s)} & \textbf{Primary Platform(s)} & \textbf{Prior XR A11y Exp} & \textbf{Evaluated} \\
\midrule
P1 & Male & startup, freelance & 5y: designer, writer, developer & Unity & Motor, Cognitive & Cognitive \\
P2 & Female & startup, freelance & 6.5y: developer & Unity & Visual, Motor & Visual \\
P3 & Male & big tech & 4y: manager, researcher, developer & BabylonJS Playground & Visual; XR tools & S\&H \\
P4 & Male & startup, freelance & 4y: manager, designer, developer & Unity & Visual, Motor & Cognitive \\
P5 & Male & big tech, freelance & 3y: designer, animator, developer & Unity & Speech, Hearing (S\&H) & Motor \\
P6 & Male & startup, freelance & 2.5y: manager, developer & Glitch & WebXR tools & Visual \\
P7 & Male & midsize & 5y: researcher, developer & Unity & Visual, Cognitive & Motor \\
P8 & Male & startup, freelance & 9y: designer, developer & Custom Engine; Unity & None & Cognitive \\
P9 & Male & midsize & 2y: developer & Glitch & None & Visual \\
P10 & Female & midsize & 3y: developer & Unity & None & S\&H \\
P11 & Male & startup, freelance & 5y: developer & Unreal & None & Motor \\
P12 & Male & startup & 3.5y: designer & Unity & Hearing & S\&H \\
P13 & Male & startup & 2y: developer & Unity & Motor & Motor \\
P14 & Male & midsize, freelance & 3y: researcher, developer & PlayCanvas & Visual & S\&H \\
P15 & Male & big tech & 5.5y: designer, developer & Unity & Visual; XR tools & Visual \\
P16 & Male & midsize, startup & 4y: designer, researcher, developer & Unity & Cognitive; XR tools & Cognitive \\
P17 & Male & midsize & <1y: a11y developer & \textit{Redacted XR Engine} & \textit{A11y team at XR Engine} & Visual \\
P18 & Male & startup & 6y: executive, designer, researcher & Unity & None & S\&H \\
P19 & Male & big tech & 8y: developer & Unity & Visual, Motor & S\&H \\
P20 & Male & startup & 7y: team lead, developer & Unity & Visual & Cognitive \\
P21 & Male & big tech, startup & 5y: developer & Unity & Motor & Cognitive \\
P22 & Male & startup, freelance & 5y: team lead, developer & Unity & Cognitive & Cognitive \\
P23 & Male & big tech, startup & 8y: designer, developer & Unity & Visual, Motor, Speech & Visual \\
P24 & Male & big tech, midsize & 7y: developer & Unity & Visual & Visual \\
P25 & Male & big tech & 5y: team lead, developer & Custom Engine; Unreal & Cognitive, S\&H & S\&H \\
\bottomrule
\end{tabular}}
\caption{Participant Demographic Information, including gender, organizational context(s), years of XR development experience and responsibilities, primary development platforms, years of XR a11y development experience, and guidelines evaluated.
}
\label{table:participants}
\end{table*}


We interviewed 25 practitioners (23 male, 2 female) with rich XR development experience. Twenty-four participants had 2--9 years experience ($mean=4.92$, $SD=1.94$); P17 had less than one year of XR development experience but worked on an a11y team at a major XR game engine. Participants represented diverse organizational contexts: freelancers ($n=7$); startups ($n=6$); midsize companies (like game engines, universities, and healthcare organizations) ($n=6$); and big tech ($n=6$), covering perspectives across different resource and constraint environments that significantly impact a11y implementation approaches~\cite{wang2025understanding}.
To ensure qualified evaluation of a11y guidelines, we pre-screened participants prioritizing those with a11y development experience and public contributions to XR projects: 20 participants had prior XR a11y experience, with most experienced in visual a11y ($n=10$), followed by motor ($n=7$), cognitive ($n=5$), speech ($n=4$), and hearing ($n=3$). Four participants had experience creating tools for XR platforms. Most participants have development experience with Unity ($n=23$), with additional experience using Unreal Engine ($n=11$) and various webXR ($n=6$) or custom/company-specific engines ($n=3$).
We detail participants' demographic information, including individuals' years of experience and preferred platform(s), in Table~\ref{table:participants}.

\subsection{Apparatus: Guideline Selection}
\label{guidelineapparatus}
We collected a11y guidelines applicable to XR through a systematic literature search combining ``Augmented Reality,'' ``Virtual Reality,'' ``Mixed Reality,'' ``Extended Reality,'' and ``Game'' with ``accessibility guidelines'' across Google and Google Scholar. While there are still no commonly-agreed, mature XR a11y standards, we collected existing preliminary guidelines distilled by researchers, a11y organizations, and \changed{industry}. \changed{Specifically, we used Google Scholar as our primary search database (as it aggregates content from major academic databases including ACM Digital Library and IEEE Xplore) to collect guidelines derived from research studies (\textit{e.g.,}~\cite{heilemann2021guidelines, w3cxr}), and we supplemented the results with a more general Google search to identify practitioner-facing guidelines from industry (\textit{e.g.,}~\cite{magicleap, xbox, oculusvr}) and a11y organizations (\textit{e.g.,}~\cite{apx, gameguide, sig2022guidelines}).} 
We included game guidelines due to significant design paradigm overlap between XR applications and 3D games~\cite{heilemann2021guidelines}. 

Our search yielded 12 resources: \cite{apx, androidaccess, gameguide, uap, heilemann2021guidelines, sig2022guidelines, magicleap, xbox, oculusvr, melbourne, w3mobile, w3cxr}. To narrow the scope we focused on relatively mature, comprehensive guideline resources, excluding guidelines that primarily provided conceptual information without specific a11y issues and implementation recommendations (e.g., \cite{melbourne}) or guidelines that were a subset of more comprehensive resources already included in our analysis (\textit{e.g.,} \cite{heilemann2021guidelines, w3mobile}). We narrowed down to six resources, including the \textit{Game Accessibility Guidelines} (GAG)~\cite{gameguide}, \textit{W3C XR Accessibility User Requirements} (W3CXR)~\cite{w3cxr}, \textit{Meta Quest Accessible VR Design} (Quest)~\cite{oculusvr}, \textit{Accessible Player Experiences} (APX)~\cite{apx}, \textit{Xbox Accessibility Guidelines} (Xbox)~\cite{xbox}, and \textit{IGDA GASIG Top Ten} (IGDA)~\cite{sig2022guidelines}. 

\changed{
\begin{figure}[h!]
    \centering
    \includegraphics[width=1\linewidth]{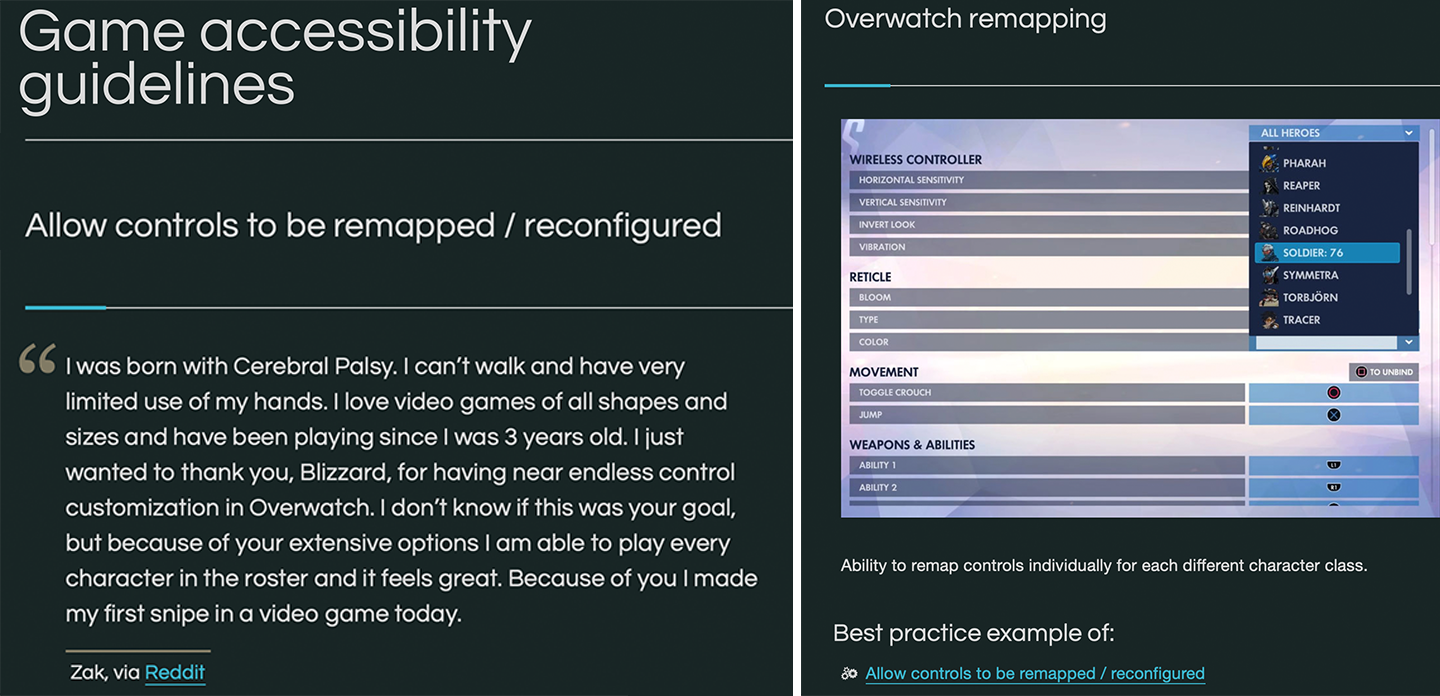}
    \caption[]{Screenshots as shown to participants from the Game Accessibility Guidelines used as an anchor, showing \textit{Mot-1: Allow controls to be remapped / reconfigured} as an example. We provided original wording (left) and visual examples (right) to participants and gave standardized explanations of each guideline via a slides presentation. Retrieved from~\cite{mot1-gameguide}.} 
    \label{fig:mot1}
    \Description{Screenshot from Game Accessibility Guidelines website showing the guideline ``Allow controls to be remapped/reconfigured'' alongside an example implementation from the game Overwatch showing its controller remapping interface. Players can modify their controls independently based on preferred control scheme and per character they want to play.}
\end{figure}
}

\aptLtoX{\begin{table*}[ht]
\footnotesize
\Description{A table presenting selected XR Accessibility Guidelines, categorized by disability type. The table has five columns: ``Disabilities'', ``Guideline ID,'' ``Guidelines'', ``References'', and ``Short Guideline''. The Disabilities column lists four categories: Visual Impairments, Motor Impairments, Cognitive Impairments, and Speech and Hearing Impairments. The Guideline ID column shows abbreviated identifiers. The Guidelines column provides detailed guidelines with source references in brackets. The References column lists source links. The Short Guideline column provides concise descriptive names.}
\centering
\begin{tabular}{\textwidth}{p{1.4cm}p{2.8cm}p{8cm}p{4.2cm}}
  \toprule
  \textbf{Disabilities} & \textbf{Guideline ID} & \textbf{Guideline Content} & \textbf{References}\\
  \toprule
  \multirow{5}{5em}{\textbf{Motor Impairments}}
 & \textit{Mot-1:} Remap Controls & Allow controls to be remapped / reconfigured. \tiny{(See Fig.~\ref{fig:mot1})} & \href{http://gameaccessibilityguidelines.com/allow-controls-to-be-remapped-reconfigured}{GAG}, \href{https://accessible.games/accessible-player-experiences/access-patterns/same-controls-but-different/}{APX}, \href{http://igda-gasig.org/get-involved/sig-initiatives/resources-for-game-developers/sig-guidelines/}{IGDA}, \href{https://developer.oculus.com/resources/design-accessible-vr-controls/}{Quest}, \href{https://www.w3.org/TR/xaur/\#xr-controller-challenges}{W3CXR}, \href{https://learn.microsoft.com/en-us/gaming/accessibility/Xbox-accessibility-guidelines/107}{Xbox}\\
 \cmidrule{2-4}
 & \textit{Mot-2:} Flexible Inputs & Support more than one input device. & \href{http://gameaccessibilityguidelines.com/support-more-than-one-input-device}{GAG}, \href{https://accessible.games/accessible-player-experiences/access-patterns/flexible-controllers/}{APX}, \href{http://igda-gasig.org/get-involved/sig-initiatives/resources-for-game-developers/sig-guidelines/}{IGDA}, \href{https://www.w3.org/TR/xaur/\#various-input-modalities}{W3CXR}, \href{https://learn.microsoft.com/en-us/gaming/accessibility/Xbox-accessibility-guidelines/107}{Xbox}\\
 \cmidrule{2-4}
 & \textit{Mot-3:} Body-Agnostic & Do not rely on motion tracking of specific body types. & \href{https://gameaccessibilityguidelines.com/do-not-rely-on-motion-tracking-of-specific-body-types/}{GAG}, \href{https://accessible.games/accessible-player-experiences/access-patterns/flexible-controllers/}{APX}, \href{https://developer.oculus.com/resources/design-accessible-vr-controls/}{Quest}, \href{https://www.w3.org/TR/xaur/\#motion-agnostic-interactions}{W3CXR}, \href{https://learn.microsoft.com/en-us/gaming/accessibility/Xbox-accessibility-guidelines/107}{Xbox}\\
 \cmidrule{2-4}
 & \textit{Mot-4:} Simple Controls & Ensure controls are as simple as possible, or provide a simpler alternative. & \href{http://gameaccessibilityguidelines.com/ensure-controls-are-as-simple-as-possible-or-provide-a-simpler-alternative}{GAG}, \href{https://accessible.games/accessible-player-experiences/access-patterns/do-more-with-less/}{APX}, \href{https://igda-gasig.org/get-involved/sig-initiatives/resources-for-game-developers/sig-guidelines/on-mobility-disabilities/}{IGDA}, \href{\detokenize{https://developer.oculus.com/resources/design-accessible-vr/\#minimize-the-complexity-of-your-controller-scheme}}{Quest}, \href{https://www.w3.org/TR/xaur/\#interaction-and-target-customization}{W3CXR}, \href{https://learn.microsoft.com/en-us/gaming/accessibility/Xbox-accessibility-guidelines/107}{Xbox}\\
 \cmidrule{2-4}
 & \textit{Mot-5:} Flexible Timing & Do not make precise timing essential to gameplay--offer alternatives, actions that can be carried out while paused, or a skip mechanism. & \href{http://gameaccessibilityguidelines.com/do-not-make-precise-timing-essential-to-gameplay-offer-alternatives-actions-that-can-be-carried-out-while-paused-or-a-skip-mechanism}{GAG}, \href{https://accessible.games/accessible-player-experiences/access-patterns/improved-precision/}{APX}, \href{https://igda-gasig.org/get-involved/sig-initiatives/resources-for-game-developers/sig-guidelines/on-mobility-disabilities/}{IGDA}, \href{https://www.w3.org/TR/xaur/\#interaction-speed}{W3CXR}, \href{https://learn.microsoft.com/en-us/gaming/accessibility/Xbox-accessibility-guidelines/116}{Xbox}\\
  \midrule
  \multirow{5}{5em}{\textbf{Visual Impairments}} 
 & \textit{Vis-1:} Supplement Color & Ensure no essential information is conveyed by a fixed colour alone. & \href{https://gameaccessibilityguidelines.com/ensure-no-essential-information-is-conveyed-by-a-colour-alone/}{GAG}, \href{https://accessible.games/accessible-player-experiences/access-patterns/distinguish-this-from-that/}{APX}, \href{https://igda-gasig.org/get-involved/sig-initiatives/resources-for-game-developers/sig-guidelines/on-visual-disabilities/}{IGDA}, \href{https://developer.oculus.com/resources/design-accessible-vr-display/\#color}{Quest}, \href{https://learn.microsoft.com/en-us/gaming/accessibility/Xbox-accessibility-guidelines/103}{Xbox}\\
 \cmidrule{2-4}
 & \textit{Vis-2:} Resizable UI & Allow interfaces to be resized. & \href{http://gameaccessibilityguidelines.com/allow-interfaces-to-be-resized}{GAG}, \href{https://accessible.games/accessible-player-experiences/access-patterns/personal-interface/}{APX}, \href{https://igda-gasig.org/get-involved/sig-initiatives/resources-for-game-developers/sig-guidelines/on-visual-disabilities/}{IGDA}, \href{https://www.w3.org/TR/xaur/\#interaction-and-target-customization}{W3CXR}, \href{https://learn.microsoft.com/en-us/gaming/accessibility/Xbox-accessibility-guidelines/112}{Xbox}\\
 \cmidrule{2-4}
 & \textit{Vis-3:} Readable Font Size & Use an easily readable default font size. & \href{http://gameaccessibilityguidelines.com/use-an-easily-readable-default-font-size}{GAG}, \href{https://accessible.games/accessible-player-experiences/access-patterns/clear-text/}{APX}, \href{https://developer.oculus.com/resources/design-accessible-vr-captions/}{Quest}, \href{https://learn.microsoft.com/en-us/gaming/accessibility/Xbox-accessibility-guidelines/101}{Xbox}\\
 \cmidrule{2-4}
 & \textit{Vis-4:} Audio Description & Provide an audio description track. & \href{http://gameaccessibilityguidelines.com/provide-an-audio-description-track/}{GAG}, \href{https://accessible.games/accessible-player-experiences/access-patterns/second_channel/}{APX}, \href{https://www.w3.org/TR/xaur/\#gestural-interfaces-and-interactions}{W3CXR}, \href{https://learn.microsoft.com/en-us/gaming/accessibility/Xbox-accessibility-guidelines/111}{Xbox}\\
 \cmidrule{2-4}
 & \textit{Vis-5:} Text Voiceovers & Provide pre-recorded voiceovers for all text, including menus and installers. & \href{http://gameaccessibilityguidelines.com/provide-full-internal-self-voicing-for-all-text-including-menus-and-installers}{GAG}, \href{https://igda-gasig.org/get-involved/sig-initiatives/resources-for-game-developers/sig-guidelines/on-visual-disabilities/}{IGDA}, \href{https://developer.oculus.com/resources/design-accessible-vr-design/}{Quest}, \href{https://www.w3.org/TR/xaur/\#immersive-semantics-and-customization}{W3CXR}, \href{https://learn.microsoft.com/en-us/gaming/accessibility/Xbox-accessibility-guidelines/106}{Xbox}\\
  \midrule
  \multirow{5}{5em}{\textbf{Cognitive Impairments}} 
 & \textit{Cog-1:} Avoid Flicker & Avoid flickering images and repetitive patterns. & \href{http://gameaccessibilityguidelines.com/avoid-flickering-images-and-repetitive-patterns}{GAG}, \href{https://accessible.games/accessible-player-experiences/access-patterns/clear-channels/}{APX}, \href{https://developer.oculus.com/resources/design-accessible-vr-ui-ux/}{Quest}, \href{https://www.w3.org/TR/xaur/\#avoiding-sickness-triggers}{W3CXR}, \href{https://learn.microsoft.com/en-us/gaming/accessibility/Xbox-accessibility-guidelines/118}{Xbox}\\
 \cmidrule{2-4}
 & \textit{Cog-2:} Supplement Text & Ensure no essential information (especially instructions) is conveyed by text alone, reinforce with visuals and/or speech. & \href{http://gameaccessibilityguidelines.com/ensure-no-essential-information-especially-instructions-is-conveyed-by-text-alone-reinforce-with-visuals-andor-speech}{GAG}, \href{https://accessible.games/accessible-player-experiences/access-patterns/second_channel/}{APX}, \href{https://igda-gasig.org/get-involved/sig-initiatives/resources-for-game-developers/sig-guidelines/on-cognitive-disabilities/}{IGDA}, \href{https://developer.oculus.com/resources/design-accessible-vr-captions/}{Quest}, \href{https://www.w3.org/TR/xaur/\#critical-messaging-and-alerts}{W3CXR}, \href{https://learn.microsoft.com/en-us/gaming/accessibility/Xbox-accessibility-guidelines/103}{Xbox}\\
 \cmidrule{2-4}
 & \textit{Cog-3:} Symbol Chat & Use symbol-based chat (smileys, etc.). & \href{http://gameaccessibilityguidelines.com/use-symbol-based-chat-smileys-etc}{GAG}, \href{https://accessible.games/accessible-player-experiences/access-patterns/flexible-text-entry/}{APX}, \href{https://igda-gasig.org/get-involved/sig-initiatives/resources-for-game-developers/sig-guidelines/on-cognitive-disabilities/}{IGDA}, \href{https://www.w3.org/TR/xaur/\#immersive-personalization}{W3CXR}, \href{https://learn.microsoft.com/en-us/gaming/accessibility/Xbox-accessibility-guidelines/120}{Xbox}\\
 \cmidrule{2-4}
 & \textit{Cog-4:} Adjustable Speed & Include an option to adjust the game speed. & \href{http://gameaccessibilityguidelines.com/include-an-option-to-adjust-the-game-speed}{GAG}, \href{https://accessible.games/accessible-player-experiences/challenge-patterns/slow-it-down/}{APX}, \href{http://igda-gasig.org/get-involved/sig-initiatives/resources-for-game-developers/sig-guidelines/}{IGDA}, \href{https://www.w3.org/TR/xaur/\#interaction-speed}{W3CXR}, \href{https://learn.microsoft.com/en-us/gaming/accessibility/Xbox-accessibility-guidelines/116}{Xbox}\\
 \cmidrule{2-4}
 & \textit{Cog-5:} Hide Distractions & Provide an option to turn off / hide background movement. & \href{http://gameaccessibilityguidelines.com/provide-an-option-to-turn-off-hide-background-movement}{GAG}, \href{https://accessible.games/accessible-player-experiences/access-patterns/distinguish-this-from-that/}{APX}, \href{http://igda-gasig.org/get-involved/sig-initiatives/resources-for-game-developers/sig-guidelines/}{IGDA}, \href{https://www.w3.org/TR/xaur/\#immersive-personalization}{W3CXR}, \href{https://learn.microsoft.com/en-us/gaming/accessibility/Xbox-accessibility-guidelines/117}{Xbox}\\
 \midrule
 \multirow{5}{5em}{\textbf{Speech \& Hearing Impairments}}
 & \textit{SH-1:} Visualize Sound & Provide captions or visuals for significant background sounds. & \href{http://gameaccessibilityguidelines.com/provide-captions-or-visuals-for-significant-background-sounds/}{GAG}, \href{https://accessible.games/accessible-player-experiences/access-patterns/second_channel/}{APX}, \href{https://igda-gasig.org/get-involved/sig-initiatives/resources-for-game-developers/sig-guidelines/on-auditory-disabilities/}{IGDA}, \href{https://developer.oculus.com/resources/design-accessible-vr-captions/}{Quest}, \href{https://www.w3.org/TR/xaur/\#spatial-audio-tracks-and-alternatives}{W3CXR}, \href{https://learn.microsoft.com/en-us/gaming/accessibility/Xbox-accessibility-guidelines/104}{Xbox}\\
 \cmidrule{2-4}
 & \textit{SH-2:} Multimodal Chat & Support text chat as well as voice for multiplayer. & \href{http://gameaccessibilityguidelines.com/support-text-chat-as-well-as-voice-for-multiplayer}{GAG}, \href{https://accessible.games/accessible-player-experiences/access-patterns/flexible-text-entry/}{APX}, \href{https://igda-gasig.org/get-involved/sig-initiatives/resources-for-game-developers/sig-guidelines/on-auditory-disabilities/}{IGDA}, \href{https://www.w3.org/TR/xaur/\#various-input-modalities}{W3CXR}, \href{https://learn.microsoft.com/en-us/gaming/accessibility/Xbox-accessibility-guidelines/120}{Xbox}\\
 \cmidrule{2-4}
 & \textit{SH-3:} Supplement Audio & Ensure no essential information is conveyed by sounds alone. & \href{http://gameaccessibilityguidelines.com/ensure-no-essential-information-is-conveyed-by-sounds-alone}{GAG}, \href{https://accessible.games/accessible-player-experiences/access-patterns/second_channel/}{APX}, \href{https://igda-gasig.org/get-involved/sig-initiatives/resources-for-game-developers/sig-guidelines/on-auditory-disabilities/}{IGDA}, \href{https://developer.oculus.com/resources/design-accessible-vr-design/}{Quest}, \href{https://www.w3.org/TR/xaur/\#xr-and-supporting-multimodality}{W3CXR}, \href{https://learn.microsoft.com/en-us/gaming/accessibility/Xbox-accessibility-guidelines/103}{Xbox}\\
 \cmidrule{2-4}
 & \textit{SH-4:} Separate Volumes & Provide separate volume controls or mutes for effects, speech and background / music. & \href{http://gameaccessibilityguidelines.com/provide-separate-volume-controls-or-mutes-for-effects-speech-and-background-music}{GAG}, \href{https://accessible.games/accessible-player-experiences/access-patterns/clear-channels/}{APX}, \href{https://igda-gasig.org/get-involved/sig-initiatives/resources-for-game-developers/sig-guidelines/on-auditory-disabilities/}{IGDA}, \href{https://developer.oculus.com/resources/design-accessible-vr-audio/}{Quest}, \href{https://www.w3.org/TR/xaur/\#immersive-personalization}{W3CXR}, \href{https://learn.microsoft.com/en-us/gaming/accessibility/Xbox-accessibility-guidelines/105}{Xbox}\\
 \cmidrule{2-4}
 & \textit{SH-5:} Subtitle Settings & Allow subtitle/caption presentation to be customised. & \href{http://gameaccessibilityguidelines.com/allow-subtitlecaption-presentation-to-be-customised/}{GAG}, \href{https://accessible.games/accessible-player-experiences/access-patterns/clear-text/}{APX}, \href{https://igda-gasig.org/get-involved/sig-initiatives/resources-for-game-developers/sig-guidelines/on-auditory-disabilities/}{IGDA}, \href{https://developer.oculus.com/resources/design-accessible-vr-captions/}{Quest}, \href{https://www.w3.org/TR/xaur/\#captioning-subtitling-and-text-support-and-customization}{W3CXR}\\
  \bottomrule
\end{tabular}
\caption[]{\changed{Selected XR A11y Guidelines, grouped by disability type (Motor, Visual, Cognitive, and Speech \& Hearing). Guidelines are abbreviated with unique IDs and short reference names for identification under the \textit{Guideline ID} column. The \textit{Guideline Content} presents the original wording of the guideline summary from the GAG resource. Original wording for each guideline from different sources are hyperlinked under the \textit{References} column (note that similar guideline content could be presented differently in different resources)}. 
}
\label{table:guidelines}
\end{table*}
}{\begin{table*}[ht]
\footnotesize
\Description{A table presenting selected XR Accessibility Guidelines, categorized by disability type. The table has five columns: ``Disabilities'', ``Guideline ID,'' ``Guidelines'', ``References'', and ``Short Guideline''. The Disabilities column lists four categories: Visual Impairments, Motor Impairments, Cognitive Impairments, and Speech and Hearing Impairments. The Guideline ID column shows abbreviated identifiers. The Guidelines column provides detailed guidelines with source references in brackets. The References column lists source links. The Short Guideline column provides concise descriptive names.}
\centering
\begin{tabular}{p{1.4cm}p{2.8cm}p{8cm}p{4.2cm}}
  \toprule
  \textbf{Disabilities} & \textbf{Guideline ID} & \textbf{Guideline Content} & \textbf{References}\\
  \toprule
  \multirow{5}{5em}{\textbf{Motor \\ Impairments}}
 & \textit{Mot-1:} Remap Controls & Allow controls to be remapped / reconfigured. \tiny{(See Fig.~\ref{fig:mot1})} & \href{http://gameaccessibilityguidelines.com/allow-controls-to-be-remapped-reconfigured}{GAG}, \href{https://accessible.games/accessible-player-experiences/access-patterns/same-controls-but-different/}{APX}, \href{http://igda-gasig.org/get-involved/sig-initiatives/resources-for-game-developers/sig-guidelines/}{IGDA}, \href{https://developer.oculus.com/resources/design-accessible-vr-controls/}{Quest}, \href{https://www.w3.org/TR/xaur/\#xr-controller-challenges}{W3CXR}, \href{https://learn.microsoft.com/en-us/gaming/accessibility/Xbox-accessibility-guidelines/107}{Xbox}\\
 \arrayrulecolor{gray}\cmidrule{2-4}\arrayrulecolor{black}
 & \textit{Mot-2:} Flexible Inputs & Support more than one input device. & \href{http://gameaccessibilityguidelines.com/support-more-than-one-input-device}{GAG}, \href{https://accessible.games/accessible-player-experiences/access-patterns/flexible-controllers/}{APX}, \href{http://igda-gasig.org/get-involved/sig-initiatives/resources-for-game-developers/sig-guidelines/}{IGDA}, \href{https://www.w3.org/TR/xaur/\#various-input-modalities}{W3CXR}, \href{https://learn.microsoft.com/en-us/gaming/accessibility/Xbox-accessibility-guidelines/107}{Xbox}\\
 \arrayrulecolor{gray}\cmidrule{2-4}\arrayrulecolor{black}
 & \textit{Mot-3:} Body-Agnostic & Do not rely on motion tracking of specific body types. & \href{https://gameaccessibilityguidelines.com/do-not-rely-on-motion-tracking-of-specific-body-types/}{GAG}, \href{https://accessible.games/accessible-player-experiences/access-patterns/flexible-controllers/}{APX}, \href{https://developer.oculus.com/resources/design-accessible-vr-controls/}{Quest}, \href{https://www.w3.org/TR/xaur/\#motion-agnostic-interactions}{W3CXR}, \href{https://learn.microsoft.com/en-us/gaming/accessibility/Xbox-accessibility-guidelines/107}{Xbox}\\
 \arrayrulecolor{gray}\cmidrule{2-4}\arrayrulecolor{black}
 & \textit{Mot-4:} Simple Controls & Ensure controls are as simple as possible, or provide a simpler alternative. & \href{http://gameaccessibilityguidelines.com/ensure-controls-are-as-simple-as-possible-or-provide-a-simpler-alternative}{GAG}, \href{https://accessible.games/accessible-player-experiences/access-patterns/do-more-with-less/}{APX}, \href{https://igda-gasig.org/get-involved/sig-initiatives/resources-for-game-developers/sig-guidelines/on-mobility-disabilities/}{IGDA}, \href{\detokenize{https://developer.oculus.com/resources/design-accessible-vr/\#minimize-the-complexity-of-your-controller-scheme}}{Quest}, \href{https://www.w3.org/TR/xaur/\#interaction-and-target-customization}{W3CXR}, \href{https://learn.microsoft.com/en-us/gaming/accessibility/Xbox-accessibility-guidelines/107}{Xbox}\\
 \arrayrulecolor{gray}\cmidrule{2-4}\arrayrulecolor{black}
 & \textit{Mot-5:} Flexible Timing & Do not make precise timing essential to gameplay--offer alternatives, actions that can be carried out while paused, or a skip mechanism. & \href{http://gameaccessibilityguidelines.com/do-not-make-precise-timing-essential-to-gameplay-offer-alternatives-actions-that-can-be-carried-out-while-paused-or-a-skip-mechanism}{GAG}, \href{https://accessible.games/accessible-player-experiences/access-patterns/improved-precision/}{APX}, \href{https://igda-gasig.org/get-involved/sig-initiatives/resources-for-game-developers/sig-guidelines/on-mobility-disabilities/}{IGDA}, \href{https://www.w3.org/TR/xaur/\#interaction-speed}{W3CXR}, \href{https://learn.microsoft.com/en-us/gaming/accessibility/Xbox-accessibility-guidelines/116}{Xbox}\\
  \midrule
  \multirow{5}{5em}{\textbf{Visual \\ Impairments}} 
 & \textit{Vis-1:} Supplement Color & Ensure no essential information is conveyed by a fixed colour alone. & \href{https://gameaccessibilityguidelines.com/ensure-no-essential-information-is-conveyed-by-a-colour-alone/}{GAG}, \href{https://accessible.games/accessible-player-experiences/access-patterns/distinguish-this-from-that/}{APX}, \href{https://igda-gasig.org/get-involved/sig-initiatives/resources-for-game-developers/sig-guidelines/on-visual-disabilities/}{IGDA}, \href{https://developer.oculus.com/resources/design-accessible-vr-display/\#color}{Quest}, \href{https://learn.microsoft.com/en-us/gaming/accessibility/Xbox-accessibility-guidelines/103}{Xbox}\\
 \arrayrulecolor{gray}\cmidrule{2-4}\arrayrulecolor{black}
 & \textit{Vis-2:} Resizable UI & Allow interfaces to be resized. & \href{http://gameaccessibilityguidelines.com/allow-interfaces-to-be-resized}{GAG}, \href{https://accessible.games/accessible-player-experiences/access-patterns/personal-interface/}{APX}, \href{https://igda-gasig.org/get-involved/sig-initiatives/resources-for-game-developers/sig-guidelines/on-visual-disabilities/}{IGDA}, \href{https://www.w3.org/TR/xaur/\#interaction-and-target-customization}{W3CXR}, \href{https://learn.microsoft.com/en-us/gaming/accessibility/Xbox-accessibility-guidelines/112}{Xbox}\\
 \arrayrulecolor{gray}\cmidrule{2-4}\arrayrulecolor{black}
 & \textit{Vis-3:} Readable Font Size & Use an easily readable default font size. & \href{http://gameaccessibilityguidelines.com/use-an-easily-readable-default-font-size}{GAG}, \href{https://accessible.games/accessible-player-experiences/access-patterns/clear-text/}{APX}, \href{https://developer.oculus.com/resources/design-accessible-vr-captions/}{Quest}, \href{https://learn.microsoft.com/en-us/gaming/accessibility/Xbox-accessibility-guidelines/101}{Xbox}\\
 \arrayrulecolor{gray}\cmidrule{2-4}\arrayrulecolor{black}
 & \textit{Vis-4:} Audio Description & Provide an audio description track. & \href{http://gameaccessibilityguidelines.com/provide-an-audio-description-track/}{GAG}, \href{https://accessible.games/accessible-player-experiences/access-patterns/second_channel/}{APX}, \href{https://www.w3.org/TR/xaur/\#gestural-interfaces-and-interactions}{W3CXR}, \href{https://learn.microsoft.com/en-us/gaming/accessibility/Xbox-accessibility-guidelines/111}{Xbox}\\
 \arrayrulecolor{gray}\cmidrule{2-4}\arrayrulecolor{black}
 & \textit{Vis-5:} Text Voiceovers & Provide pre-recorded voiceovers for all text, including menus and installers. & \href{http://gameaccessibilityguidelines.com/provide-full-internal-self-voicing-for-all-text-including-menus-and-installers}{GAG}, \href{https://igda-gasig.org/get-involved/sig-initiatives/resources-for-game-developers/sig-guidelines/on-visual-disabilities/}{IGDA}, \href{https://developer.oculus.com/resources/design-accessible-vr-design/}{Quest}, \href{https://www.w3.org/TR/xaur/\#immersive-semantics-and-customization}{W3CXR}, \href{https://learn.microsoft.com/en-us/gaming/accessibility/Xbox-accessibility-guidelines/106}{Xbox}\\
  \midrule
  \multirow{5}{5em}{\textbf{Cognitive \\ Impairments}} 
 & \textit{Cog-1:} Avoid Flicker & Avoid flickering images and repetitive patterns. & \href{http://gameaccessibilityguidelines.com/avoid-flickering-images-and-repetitive-patterns}{GAG}, \href{https://accessible.games/accessible-player-experiences/access-patterns/clear-channels/}{APX}, \href{https://developer.oculus.com/resources/design-accessible-vr-ui-ux/}{Quest}, \href{https://www.w3.org/TR/xaur/\#avoiding-sickness-triggers}{W3CXR}, \href{https://learn.microsoft.com/en-us/gaming/accessibility/Xbox-accessibility-guidelines/118}{Xbox}\\
 \arrayrulecolor{gray}\cmidrule{2-4}\arrayrulecolor{black}
 & \textit{Cog-2:} Supplement Text & Ensure no essential information (especially instructions) is conveyed by text alone, reinforce with visuals and/or speech. & \href{http://gameaccessibilityguidelines.com/ensure-no-essential-information-especially-instructions-is-conveyed-by-text-alone-reinforce-with-visuals-andor-speech}{GAG}, \href{https://accessible.games/accessible-player-experiences/access-patterns/second_channel/}{APX}, \href{https://igda-gasig.org/get-involved/sig-initiatives/resources-for-game-developers/sig-guidelines/on-cognitive-disabilities/}{IGDA}, \href{https://developer.oculus.com/resources/design-accessible-vr-captions/}{Quest}, \href{https://www.w3.org/TR/xaur/\#critical-messaging-and-alerts}{W3CXR}, \href{https://learn.microsoft.com/en-us/gaming/accessibility/Xbox-accessibility-guidelines/103}{Xbox}\\
 \arrayrulecolor{gray}\cmidrule{2-4}\arrayrulecolor{black}
 & \textit{Cog-3:} Symbol Chat & Use symbol-based chat (smileys, etc.). & \href{http://gameaccessibilityguidelines.com/use-symbol-based-chat-smileys-etc}{GAG}, \href{https://accessible.games/accessible-player-experiences/access-patterns/flexible-text-entry/}{APX}, \href{https://igda-gasig.org/get-involved/sig-initiatives/resources-for-game-developers/sig-guidelines/on-cognitive-disabilities/}{IGDA}, \href{https://www.w3.org/TR/xaur/\#immersive-personalization}{W3CXR}, \href{https://learn.microsoft.com/en-us/gaming/accessibility/Xbox-accessibility-guidelines/120}{Xbox}\\
 \arrayrulecolor{gray}\cmidrule{2-4}\arrayrulecolor{black}
 & \textit{Cog-4:} Adjustable Speed & Include an option to adjust the game speed. & \href{http://gameaccessibilityguidelines.com/include-an-option-to-adjust-the-game-speed}{GAG}, \href{https://accessible.games/accessible-player-experiences/challenge-patterns/slow-it-down/}{APX}, \href{http://igda-gasig.org/get-involved/sig-initiatives/resources-for-game-developers/sig-guidelines/}{IGDA}, \href{https://www.w3.org/TR/xaur/\#interaction-speed}{W3CXR}, \href{https://learn.microsoft.com/en-us/gaming/accessibility/Xbox-accessibility-guidelines/116}{Xbox}\\
 \arrayrulecolor{gray}\cmidrule{2-4}\arrayrulecolor{black}
 & \textit{Cog-5:} Hide Distractions & Provide an option to turn off / hide background movement. & \href{http://gameaccessibilityguidelines.com/provide-an-option-to-turn-off-hide-background-movement}{GAG}, \href{https://accessible.games/accessible-player-experiences/access-patterns/distinguish-this-from-that/}{APX}, \href{http://igda-gasig.org/get-involved/sig-initiatives/resources-for-game-developers/sig-guidelines/}{IGDA}, \href{https://www.w3.org/TR/xaur/\#immersive-personalization}{W3CXR}, \href{https://learn.microsoft.com/en-us/gaming/accessibility/Xbox-accessibility-guidelines/117}{Xbox}\\
 \midrule
 \multirow{5}{5em}{\textbf{Speech \& Hearing \\ Impairments}}
 & \textit{SH-1:} Visualize Sound & Provide captions or visuals for significant background sounds. & \href{http://gameaccessibilityguidelines.com/provide-captions-or-visuals-for-significant-background-sounds/}{GAG}, \href{https://accessible.games/accessible-player-experiences/access-patterns/second_channel/}{APX}, \href{https://igda-gasig.org/get-involved/sig-initiatives/resources-for-game-developers/sig-guidelines/on-auditory-disabilities/}{IGDA}, \href{https://developer.oculus.com/resources/design-accessible-vr-captions/}{Quest}, \href{https://www.w3.org/TR/xaur/\#spatial-audio-tracks-and-alternatives}{W3CXR}, \href{https://learn.microsoft.com/en-us/gaming/accessibility/Xbox-accessibility-guidelines/104}{Xbox}\\
 \arrayrulecolor{gray}\cmidrule{2-4}\arrayrulecolor{black}
 & \textit{SH-2:} Multimodal Chat & Support text chat as well as voice for multiplayer. & \href{http://gameaccessibilityguidelines.com/support-text-chat-as-well-as-voice-for-multiplayer}{GAG}, \href{https://accessible.games/accessible-player-experiences/access-patterns/flexible-text-entry/}{APX}, \href{https://igda-gasig.org/get-involved/sig-initiatives/resources-for-game-developers/sig-guidelines/on-auditory-disabilities/}{IGDA}, \href{https://www.w3.org/TR/xaur/\#various-input-modalities}{W3CXR}, \href{https://learn.microsoft.com/en-us/gaming/accessibility/Xbox-accessibility-guidelines/120}{Xbox}\\
 \arrayrulecolor{gray}\cmidrule{2-4}\arrayrulecolor{black}
 & \textit{SH-3:} Supplement Audio & Ensure no essential information is conveyed by sounds alone. & \href{http://gameaccessibilityguidelines.com/ensure-no-essential-information-is-conveyed-by-sounds-alone}{GAG}, \href{https://accessible.games/accessible-player-experiences/access-patterns/second_channel/}{APX}, \href{https://igda-gasig.org/get-involved/sig-initiatives/resources-for-game-developers/sig-guidelines/on-auditory-disabilities/}{IGDA}, \href{https://developer.oculus.com/resources/design-accessible-vr-design/}{Quest}, \href{https://www.w3.org/TR/xaur/\#xr-and-supporting-multimodality}{W3CXR}, \href{https://learn.microsoft.com/en-us/gaming/accessibility/Xbox-accessibility-guidelines/103}{Xbox}\\
 \arrayrulecolor{gray}\cmidrule{2-4}\arrayrulecolor{black}
 & \textit{SH-4:} Separate Volumes & Provide separate volume controls or mutes for effects, speech and background / music. & \href{http://gameaccessibilityguidelines.com/provide-separate-volume-controls-or-mutes-for-effects-speech-and-background-music}{GAG}, \href{https://accessible.games/accessible-player-experiences/access-patterns/clear-channels/}{APX}, \href{https://igda-gasig.org/get-involved/sig-initiatives/resources-for-game-developers/sig-guidelines/on-auditory-disabilities/}{IGDA}, \href{https://developer.oculus.com/resources/design-accessible-vr-audio/}{Quest}, \href{https://www.w3.org/TR/xaur/\#immersive-personalization}{W3CXR}, \href{https://learn.microsoft.com/en-us/gaming/accessibility/Xbox-accessibility-guidelines/105}{Xbox}\\
 \arrayrulecolor{gray}\cmidrule{2-4}\arrayrulecolor{black}
 & \textit{SH-5:} Subtitle Settings & Allow subtitle/caption presentation to be customised. & \href{http://gameaccessibilityguidelines.com/allow-subtitlecaption-presentation-to-be-customised/}{GAG}, \href{https://accessible.games/accessible-player-experiences/access-patterns/clear-text/}{APX}, \href{https://igda-gasig.org/get-involved/sig-initiatives/resources-for-game-developers/sig-guidelines/on-auditory-disabilities/}{IGDA}, \href{https://developer.oculus.com/resources/design-accessible-vr-captions/}{Quest}, \href{https://www.w3.org/TR/xaur/\#captioning-subtitling-and-text-support-and-customization}{W3CXR}\\
  \bottomrule
\end{tabular}
\caption[]{\changed{Selected XR A11y Guidelines, grouped by disability type (Motor, Visual, Cognitive, and Speech \& Hearing). Guidelines are abbreviated with unique IDs and short reference names for identification under the \textit{Guideline ID} column. The \textit{Guideline Content} presents the original wording of the guideline summary from the GAG resource. Original wording for each guideline from different sources are hyperlinked under the \textit{References} column (note that similar guideline content could be presented differently in different resources)}. 
}
\label{table:guidelines}
\end{table*}}

Based on these guideline resources, we sought to select a commonly-agreed subset of guidelines to present to the XR practitioners in our study, ensuring the representativeness of the guidelines and to avoid overloading the participants with too many guidelines. We compared and cross-referenced the guidelines in the six resources, organized the guidelines by disability types (visual, motor, cognitive, speech/hearing), and selected five most commonly mentioned guidelines across the six resources for each disability group. We merged the guidelines for speech and hearing disabilities as both emerged as highly relevant to communication, and they had a smaller number of guidelines than other disability types. Some resources already merged these guidelines; for example, the GAG noted text chat as benefiting d/Deaf and non-verbal users together~\cite{sh2-gameguide}. This process yielded 20 guidelines with five in each disability group, which we have aggregated in Table~\ref{table:guidelines}. 

When presenting the selected guidelines to the XR practitioners, we used GAG as our ``anchor'' resource as it 
represented the most sophisticated and comprehensive resource at the time of our study~\cite{heilemann2021guidelines}, providing elaboration, PWD experience, and best practice examples for each 
guideline (see Fig. \ref{fig:mot1}).  
During interviews, we presented participants with the actual GAG guideline wording and visual examples via a slideshow presentation as shown in Fig.~\ref{fig:mot1}, and provided links to original source materials for reference, ensuring that participants evaluated the guidelines as published.


\subsection{Procedure}
To evaluate these guidelines with XR practitioners we conducted 1.5--3 hour semi-structured interviews via Zoom (October 2022--July 2023), compensating participants at \$50/hr. Each interview included three phases:

\textit{\textbf{Background \& XR A11y Experience.}} Participants first described their demographic information, XR development experience, and typical project workflows. We then explored participants' a11y knowledge and XR-specific a11y experience. Participants with a11y experience detailed their development processes, motivations, strategies, and testing approaches. Those without XR a11y experience discussed barriers to implementation. Participants finally screen-shared one XR project to demonstrate their development practices.


\textit{\textbf{Guideline Evaluation.}} In this core phase, participants reviewed and evaluated the XR a11y guidelines we collected (\S~\ref{guidelineapparatus}). 
Due to time constraints, each participant evaluated one disability type (visual, motor, cognitive, or speech \& hearing) based on their interests and experience. We used a preference-matching approach: participants selected two preferred disability types, and we chose one to ensure balanced coverage across all guideline groups while respecting participant expertise. \textit{After cleaning the data, we resulted in seven participants evaluating visual, four motor, seven cognitive, and seven speech \& hearing guidelines. We believe we reached saturation with this count.} 
After determining the group to evaluate, we followed a structured \changed{seven}-step evaluation protocol to evaluate each guideline: \changed{(1) Researcher presents guideline using a prepared slides presentation as a visual aid, showing guidelines' original wording from GAG, published examples from source (Fig.~\ref{fig:mot1}), and links to original resources for reference;} (2) Participant provides initial interpretation of the guideline \changed{in their own words}; (3) Researcher provides standardized explanation if interpretation differs; (4) Participant assesses implementation feasibility within their specific XR development context; (5) Detailed discussion of anticipated technical challenges and implementation barriers; (6) Participant suggests adaptations or improvements to the guideline for better applying to XR contexts; (7) Participant assesses external support needed (tools, resources, expertise) for successful implementation. This protocol enabled systematic comparison across participants while capturing contextual factors affecting guideline applicability. 

\textit{\textbf{Development Support Preferences.}} We concluded by exploring participants' perspectives and preferences on a11y integration beyond guidelines (\textit{e.g.,} toolkits, \textit{post hoc} solutions) and resource allocation attitudes.

\subsection{Analysis}
\changed{Interviews were recorded with Zoom and transcribed by a professional online service approved by the University of Wisconsin--Madison Institutional Review Board (IRB). Researchers reviewed each transcript to correct errors. We analyzed transcripts using thematic analysis~\cite{braun2012thematic}, starting with two researchers open coding the same set of three participants independently, then meeting to calibrate interpretations and resolve disagreements through discussion. Through iterative discussion across multiple coding sessions, researchers ultimately agreed on all codes used for 100\% inter-coder reliability.} One researcher coded the remaining 22 transcripts, updating the codebook as new codes emerged upon agreement with the research team. 
Analysis prioritized themes revealing: (1) gaps between XR development practices and guidelines' suggestions, (2) interpretation challenges when applying 3D guidelines to XR and  implementation barriers specific to XR contexts, and (3) suggested modifications for XR-appropriate guidelines. 



\section{Findings}


\label{findings}
From interviews with 25 XR practitioners, we present key findings revealing practitioners' approaches to implementing accessible XR features, their interpretation and evaluation of existing XR-adjacent a11y guidelines, and their requirements for additional XR a11y support methods. 


\subsection{Current XR A11y Solutions and Gaps in Practice}
\label{findings-solutions}
Supporting prior work~\cite{wang2025understanding}, most participants (17 out of 25) had little to no formal XR training, instead learning through online resources ($n=14$), on the job ($n=8$), or from personal projects ($n=3$). Furthermore, no participant indicated formal a11y training. 
Despite limited resources and guidance, many XR practitioners have created innovative a11y solutions that could inform future standardization efforts, extending beyond the general implementation approaches documented by prior work. 
Our participants demonstrated unique technical approaches that reveal both successful patterns and critical gaps in current practice.

\subsubsection{Leveraging Platform Capabilities for A11y}
\label{nativedevicehooks}
Some practitioners strategically used their platform's capabilities to implement a11y features. 
For example, when developing mobile AR applications, instead of using Unity that did not ``hook'' well with the native screen reader (\textit{e.g.,} VoiceOver for iOS, Talkback for Android) (P3, P17), P3 leveraged ``Native'' development platforms (\textit{e.g.,} Babylon Native) and  
added invisible 2D buttons on top of 3D objects in their scene, so that the screen reader can access the alt text of the 2D button to read out the 3D object: \textit{``These invisible buttons, you could cycle through just as you would any other button. And they could kind of highlight with the screen reader focus indicator, ... And it would ... give you an object, or it'll basically tell you the manipulation of the 3D objects that you [can] do.''} This approach demonstrates practitioners' adaptability leveraging existing platform capabilities and choosing suitable platforms based on a11y needs.

\subsubsection{Alternative Input Integration}
Practitioners have integrated diverse input mechanisms to accommodate different a11y needs, providing practical solutions to motor guidelines (\textit{e.g., Mot-1: Remap Controls}). Participants reported using controllers like the Xbox Adaptive Controller (P20) or libraries like Rewired~\cite{rewired} (P5) and Unity's XR Input system. P17 worked at an XR game engine company and his team incorporated keyboard navigation into their XR controller system, \textit{``So any game that already had controller support now suddenly had keyboard support,}'' enabling alternate control methods for many new games developed on their engine. This solution demonstrates efficient a11y feature scaling but heavily relies on development platforms' support. 

\subsubsection{Professional Tensions in A11y Implementation} \label{sec:tensions-findings}
Nearly half of our participants ($n=12$) reported a fundamental tension between creating immersive XR experiences and implementing a11y features, extending findings from prior work on developer attitudes~\cite{wang2025understanding}. At least three practitioners (P4, P10, P20) explicitly mentioned prioritizing their game's \textit{feel} over users' a11y needs. For example, P4 hesitated to remove flickering visual effects despite awareness of seizure risks: \textit{``For some app designs, just trying to incorporate certain a11y guidelines is going to be a \textbf{cursed problem that you cannot solve}. And just to be aware of those problems at a design stage and conscientiously make that trade off, like---\textbf{we care enough} about developing this app \textbf{that we are willing to exclude} this chunk of potential players.''} This decision to exclude PWD reveals how practitioners construct professional identity through conscious choices about whose experiences matter most, pulling against their willingness to implement guidelines. 

\subsubsection{Motivating Developer Adoption} \label{sec:motivating}
Echoing prior work~\cite{microsoftInclusive101}, we found that general best design considerations for XR a11y often correlate with general best practices for XR applications ($n=10$). For example, practitioners considered their app being one-handable to be a proper design consideration (P2, P24) rather than an a11y feature. As P12 stated: \textit{``[The] best argument for a11y is usability. The fact that like, good a11y isn't just for people with impairments, it also makes the product better''}.

P6 recommended framing guidelines as \textit{``VR guidelines for everybody''} or \textit{``How to make [VR] comfortable for everyone''} rather than focusing solely on PWD to avoid XR developers ``brushing off'' guidelines' suggestions. Nearly all participants seek hard data on how many users they would gain by addressing each feature ($n=21$), the estimated time necessary to implement features ($n=19$), and approximate metrics of difficulty versus value for each feature. If a11y guidelines are framed as general design best practices, then the number of users becomes \textit{everyone} rather than framing PWD as an ``other.''

Practitioners indicated that popular apps should act as guideposts and standards for a11y implementation. P4, P21, and P24 specifically wanted guideline implementation examples that reference adoption in existing or popular apps (\textit{e.g.,} Beat Saber's colorblindness options, shown in Appendix~\ref{appendix:beatsaber}). 
When users become familiar with an a11y feature in one app, they may expect it in others, making implementation easier to justify and replicate once there's an established point of reference. 

\subsection{How Practitioners Experience and Evaluate A11y Guidelines}
\label{findings-guidelines} \label{sec:evaluation}
Our central contribution lies in understanding how practitioners interpret and evaluate existing 3D a11y guidelines when implementing their XR apps. While practitioners were familiar with \textit{some} existing a11y guidelines, they lacked experience applying them to XR apps. Even known guidelines proved insufficient for XR; for example, P3 and P20 cited Microsoft's MR Accessibility Standards~\cite{MSFTa11yMRTK3} as not ``fleshed out,'' while P3 and P17 noted that the W3C's WCAG~\cite{w3web} only applies to 2D interfaces. P24 observed that XR lacks basic system-wide a11y features like screen magnifiers despite prior research demonstrating their feasibility (\textit{e.g.,} SeeingVR's magnifier tool~\cite{zhao2019seeingvr}). Although a11y guidelines exist for traditional platforms and some 3D environments, our findings reveal significant gaps highlighting the need for accessibility guidelines specifically designed for XR's unique interaction paradigms.

\subsubsection{Practitioners' Post-Guideline Solutions}
Our systematic evaluation revealed participants' ability to translate abstract guidelines to concrete technical solutions. 
All 25 practitioners came up with a11y implementation solutions based on the guidelines that can potentially be integrated in their current projects.
For example, after evaluating \textit{SH-1: Visualize Sound}, P3 proposed an a11y-conscious technical architecture that fits into a common `good design' practice: \textit{``Say if you have a prefab that [manages] audio, like send[s] data to your visual sound indicator. So no matter, as long as you have your prefabs all extend that single singular like script or object or hook or whatever, then it should be fairly simple.''} 
P17 and P18 referenced Fortnite's radial sound indicator~\cite{delaney2021fortnite} as an existing implementation of \textit{SH-1: Visualize Sound}, while P12 innovated beyond existing solutions, describing a ``sound wheel'' objective marker that would only display when sounds occurred outside the player's field of view. For \textit{SH-5: Subtitle Settings}, P18 proposed a minimally-intrusive startup experience as simple as: \textit{``Can you see this text clearly? And if it is, click Yes,''} avoiding the need for users to navigate complex a11y menus.


\subsubsection{Confusion \& Ambiguity of the Guidelines.} \label{sec:confusion}
When evaluating our presented guidelines, practitioners generally reported understanding the three information redundancy guidelines (\textit{Vis-1: Supplement Color}, \textit{Cog-2: Supplement Text}, \textit{SH-3: Supplement Audio}) due to their similarity to established web standards. However, practitioners evaluated other guidelines as too broad or using ambiguous language; for \textit{Cog-5: Hide Distractions}, P16 and P21 struggled to define ``non-interactive elements''---does the guideline include atmospheric effects? UI chrome? Background NPCs? 
Furthermore, \textit{Vis-4: Audio Description} confused P6 and P9 about whether the guideline referred to real-time narration versus pre-recorded tracks; and how it differs from \textit{Vis-5: Text Voiceovers}, interpreting \textit{Vis-4} as inclusive of or an extension of \textit{Vis-5}. 

\subsubsection{Applicability \& Implementation Challenges in XR Contexts}
Guidelines addressing temporal flexibility \textit{(Mot-5: Flexible Timing} and \textit{Cog-4: Adjustable Speed)} presented unique implementation challenges for XR; while technically possible, participants questioned their applicability to real-time XR experiences: 
P20's firefighting simulator requires real-time reactions for training validity; P1, P8, and P13 referenced Beat Saber~\cite{beatsaber} reducing scores for slower speeds as a fairness mechanism, but we note this implementation penalizes users needing the accommodation.
Participants rated these timing-related guidelines as moderate importance but low motivation to implement as a result, struggling with preserving immersion and multiplayer fairness. P8 suggested to \textit{``develop toward your extremes, low and high, and then everything else will kind of modulate in place''}, but this strategy may still prove challenging to practitioners. This tension between a11y and core gameplay mechanics appeared uniquely challenging for XR compared to traditional gaming.

\subsubsection{Priority of Guideline Implementation} \label{sec:applicability}
Priority ratings revealed three main tiers based on perceived importance and feasibility: 

\textbf{\textit{Tier 1 (immediate implementation)}} included guidelines critical for safety or legal compliance. For example: \textit{Cog-1: Avoid Flicker}, \textit{Vis-1: Supplement Color}, and \textit{SH-3: Supplement Audio}, which participants rated as both highest importance and clearest to currently understand their wording as applied to XR.

\textbf{\textit{Tier 2 (platform-dependent)}} included features participants felt platforms should provide, \textit{e.g.}, \textit{SH-5: Subtitle Settings} and \textit{Vis-2: Resizable UI}. P3 argued the platform should handle subtitle preferences, with developers just responsible for passing data through from the users' settings. Most guidelines fell in this more moderate range with medium to high importance but similarly moderate difficulty and low motivation to implement.

\textbf{\textit{Tier 3 (context-specific)}} included guidelines with applicability concerns: \textit{Cog-3: Symbol Chat} for non-social apps, or \textit{Cog-4: Adjustable Speed} for real-time experiences. These received lowest motivation scores despite recognized importance, with freelancers like P1 neutral about implementing the guideline unless their client directly requested the feature.

\aptLtoX{\begin{table*}[ht]
\centering
\begin{tabular}{@{}p{3.1cm} l c c c c l@{}}
\toprule
\textbf{Guideline ID} & \textbf{Designers} & \textbf{\hspace{-0.2em}Developers\hspace{-0.2em}} & \textbf{QA / Test} & \textbf{\hspace{-0.2em}PM / Lead\text{*}\hspace{-0.4em}} & \textbf{\hspace{-0.2em}Audio Eng\hspace{-0.2em}} & \textbf{Phase?} \\ \midrule
\textit{Mot-1:} Remap Controls & \cellcolor{colorA}A \tiny(Technical) & \cellcolor{colorR}R & \cellcolor{colorC}C & \cellcolor{colorI}I & - &\cellcolor{phaseDesign}Design \\
\textit{Mot-2:} Flexible Inputs & \cellcolor{colorR}R \tiny(Technical) & \cellcolor{colorA}A & \cellcolor{colorC}C & \cellcolor{colorI}I & - &\cellcolor{phaseEarlyDev}Early Dev \\
\textit{Mot-3:} Body-Agnostic & \cellcolor{colorA}A \tiny(Technical) & \cellcolor{colorR}R & \cellcolor{colorC}C & \cellcolor{colorI}I & - &\cellcolor{phaseDesign}Design \\
\textit{Mot-4:} Simple Controls & \cellcolor{colorR}R \tiny(UX) & \cellcolor{colorR}R & \cellcolor{colorC}C & \cellcolor{colorI}I & - &\cellcolor{phaseEarlyDev}Early Dev \\
\textit{Mot-5:} Flexible Timing & \cellcolor{colorR}R \tiny(UX) & \cellcolor{colorR}R & \cellcolor{colorC}C & \cellcolor{colorI}I & - &\cellcolor{phaseEarlyDev}Early Dev \\
\midrule
\textit{Vis-1:} Supplement Color & \cellcolor{colorR}R \tiny(Visual) & \cellcolor{colorR}R & \cellcolor{colorC}C & \cellcolor{colorI}I & - &\cellcolor{phaseDesign}Design \\
\textit{Vis-2:} Resizable UI & \cellcolor{colorR}R \tiny(UX) & \cellcolor{colorA}A & \cellcolor{colorC}C & \cellcolor{colorC}C & - &\cellcolor{phaseDesign}Design \\
\textit{Vis-3:} Readable Font Size & \cellcolor{colorR}R \tiny(Visual) & \cellcolor{colorA}A & \cellcolor{colorC}C & \cellcolor{colorI}I & - &\cellcolor{phaseEarlyDev}Early Dev \\
\textit{Vis-4:} Audio Description & \cellcolor{colorR}R \tiny(UX) & \cellcolor{colorA}A & \cellcolor{colorC}C & \cellcolor{colorC}C & \cellcolor{colorA}A & \cellcolor{phaseMidDev}Mid- Dev \\
\textit{Vis-5:} Text Voiceovers & \cellcolor{colorR}R \tiny(UX) & \cellcolor{colorA}A & \cellcolor{colorC}C & \cellcolor{colorC}C & \cellcolor{colorA}A & \cellcolor{phaseEarlyDev}Early Dev \\
\midrule
\textit{Cog-1:} Avoid Flicker & \cellcolor{colorC}C \tiny(Technical) & \cellcolor{colorA}A & \cellcolor{colorR}R & \cellcolor{colorA}A & - &\cellcolor{phaseDesign}Design \\
\textit{Cog-2:} Supplement Text & \cellcolor{colorR}R \tiny(UX) & \cellcolor{colorA}A & \cellcolor{colorC}C & \cellcolor{colorA}A & \cellcolor{colorC}C & \cellcolor{phaseEarlyDev}Early Dev \\
\textit{Cog-3:} Symbol Chat & \cellcolor{colorR}R \tiny(UX) & \cellcolor{colorA}A & \cellcolor{colorR}R & \cellcolor{colorI}I & \cellcolor{colorC}C & \cellcolor{phaseEarlyDev}Early Dev \\
\textit{Cog-4:} Adjustable Speed & \cellcolor{colorR}R \tiny(UX) & \cellcolor{colorR}R & \cellcolor{colorC}C & \cellcolor{colorI}I & - &\cellcolor{phaseMidDev}Mid- Dev \\
\textit{Cog-5:} Hide Distractions & \cellcolor{colorR}R \tiny(UX) & \cellcolor{colorA}A & \cellcolor{colorC}C & \cellcolor{colorI}I & - &\cellcolor{phasePostLaunch}Post-Launch \\
\midrule
\textit{SH-1:} Visualize Sound & \cellcolor{colorR}R \tiny(UX) & \cellcolor{colorR}R & \cellcolor{colorC}C & \cellcolor{colorC}C & \cellcolor{colorA}A & \cellcolor{phaseEarlyDev}Early Dev \\
\textit{SH-2:} Multimodal Chat & \cellcolor{colorR}R \tiny(UX) & \cellcolor{colorA}A & \cellcolor{colorC}C & \cellcolor{colorI}I & \cellcolor{colorC}C & \cellcolor{phaseEarlyDev}Early Dev \\
\textit{SH-3:} Supplement Audio & \cellcolor{colorR}R \tiny(UX) & \cellcolor{colorA}A & \cellcolor{colorC}C & \cellcolor{colorC}C & \cellcolor{colorA}A & \cellcolor{phaseEarlyDev}Early Dev \\
\textit{SH-4:} Separate Volumes & \cellcolor{colorR}R \tiny(Audio/UX) & \cellcolor{colorR}R & \cellcolor{colorC}C & \cellcolor{colorC}C & \cellcolor{colorA}A & \cellcolor{phaseMidDev}Mid- Dev \\
\textit{SH-5:} Subtitle Settings & \cellcolor{colorR}R \tiny(UX) & \cellcolor{colorA}A & \cellcolor{colorC}C & \cellcolor{colorI}I & \cellcolor{colorC}C & \cellcolor{phasePostLaunch}Post-Launch \\ \bottomrule
\end{tabular}
\caption[]{\changed{Responsibility Assignment Matrix (RACI) with Implementation Phases. Roles are color-coded based on assignment: \colorbox{colorR}{\textbf{R}esponsible}, \colorbox{colorA}{\textbf{A}ccountable}, \colorbox{colorC}{\textbf{C}onsulted}, and \colorbox{colorI}{\textbf{I}nformed}. A dash (-) indicates no involvement necessary. Designer types shown in parentheses. The \textbf{Phase?} column shows when guidelines should be implemented: \colorbox{phaseDesign}{\textbf{Design}} (app architecture), \colorbox{phaseEarlyDev}{\textbf{Early Dev}} (prototyping/MVP), \colorbox{phaseMidDev}{\textbf{Mid- Dev}} (secondary features/polish), or \colorbox{phasePostLaunch}{\textbf{Post-Launch}} (maintenance). \\
\noindent\ \text{*}\textit{PM/Lead represents the equivalent role for a project manager at each organization's respective size: }i.e., \textit{Project/Product Manager (big tech/midsize), Technical Lead/Founder (startups), or self-coordinated/client-driven (freelancers).}}}
\Description{A responsibility assignment matrix (RACI) displaying 20 accessibility guidelines grouped into four categories: Motor (5 guidelines), Visual (5 guidelines), Cognitive (5 guidelines), and Sensory/Hearing (5 guidelines). Each guideline is assigned responsibility levels to five roles—Designers, Developers, QA/Test, PM/Lead, and Audio Engineer—using color-coded designations: Red (Responsible), Yellow (Accountable), Blue (Consulted), and Gray (Informed). A final column indicates the recommended implementation phase: Design, Early Dev, Mid-Dev, or Post-Launch. For example, "Mot-1: Remap Controls" assigns Designers as Accountable, Developers as Responsible, QA as Consulted, and PM as Informed, with implementation during the Design phase.}
\label{tab:raci}
\end{table*}}{\begin{table*}[ht]
\centering
\begin{tabular}{@{}p{3.1cm} >{\columncolor{white}}l >{\columncolor{white}}c >{\columncolor{white}}c >{\columncolor{white}}c >{\columncolor{white}}c >{\columncolor{white}}l@{}}
\toprule
\textbf{Guideline ID} & \textbf{Designers} & \textbf{\hspace{-0.2em}Developers\hspace{-0.2em}} & \textbf{QA / Test} & \textbf{\hspace{-0.2em}PM / Lead\text{*}\hspace{-0.4em}} & \textbf{\hspace{-0.2em}Audio Eng\hspace{-0.2em}} & \textbf{Phase?} \\ \midrule
\textit{Mot-1:} Remap Controls & \cellcolor{colorA}A \tiny(Technical) & \cellcolor{colorR}R & \cellcolor{colorC}C & \cellcolor{colorI}I & - &\cellcolor{phaseDesign}Design \\
\textit{Mot-2:} Flexible Inputs & \cellcolor{colorR}R \tiny(Technical) & \cellcolor{colorA}A & \cellcolor{colorC}C & \cellcolor{colorI}I & - &\cellcolor{phaseEarlyDev}Early Dev \\
\textit{Mot-3:} Body-Agnostic & \cellcolor{colorA}A \tiny(Technical) & \cellcolor{colorR}R & \cellcolor{colorC}C & \cellcolor{colorI}I & - &\cellcolor{phaseDesign}Design \\
\textit{Mot-4:} Simple Controls & \cellcolor{colorR}R \tiny(UX) & \cellcolor{colorR}R & \cellcolor{colorC}C & \cellcolor{colorI}I & - &\cellcolor{phaseEarlyDev}Early Dev \\
\textit{Mot-5:} Flexible Timing & \cellcolor{colorR}R \tiny(UX) & \cellcolor{colorR}R & \cellcolor{colorC}C & \cellcolor{colorI}I & - &\cellcolor{phaseEarlyDev}Early Dev \\
\midrule
\textit{Vis-1:} Supplement Color & \cellcolor{colorR}R \tiny(Visual) & \cellcolor{colorR}R & \cellcolor{colorC}C & \cellcolor{colorI}I & - &\cellcolor{phaseDesign}Design \\
\textit{Vis-2:} Resizable UI & \cellcolor{colorR}R \tiny(UX) & \cellcolor{colorA}A & \cellcolor{colorC}C & \cellcolor{colorC}C & - &\cellcolor{phaseDesign}Design \\
\textit{Vis-3:} Readable Font Size & \cellcolor{colorR}R \tiny(Visual) & \cellcolor{colorA}A & \cellcolor{colorC}C & \cellcolor{colorI}I & - &\cellcolor{phaseEarlyDev}Early Dev \\
\textit{Vis-4:} Audio Description & \cellcolor{colorR}R \tiny(UX) & \cellcolor{colorA}A & \cellcolor{colorC}C & \cellcolor{colorC}C & \cellcolor{colorA}A & \cellcolor{phaseMidDev}Mid- Dev \\
\textit{Vis-5:} Text Voiceovers & \cellcolor{colorR}R \tiny(UX) & \cellcolor{colorA}A & \cellcolor{colorC}C & \cellcolor{colorC}C & \cellcolor{colorA}A & \cellcolor{phaseEarlyDev}Early Dev \\
\midrule
\textit{Cog-1:} Avoid Flicker & \cellcolor{colorC}C \tiny(Technical) & \cellcolor{colorA}A & \cellcolor{colorR}R & \cellcolor{colorA}A & - &\cellcolor{phaseDesign}Design \\
\textit{Cog-2:} Supplement Text & \cellcolor{colorR}R \tiny(UX) & \cellcolor{colorA}A & \cellcolor{colorC}C & \cellcolor{colorA}A & \cellcolor{colorC}C & \cellcolor{phaseEarlyDev}Early Dev \\
\textit{Cog-3:} Symbol Chat & \cellcolor{colorR}R \tiny(UX) & \cellcolor{colorA}A & \cellcolor{colorR}R & \cellcolor{colorI}I & \cellcolor{colorC}C & \cellcolor{phaseEarlyDev}Early Dev \\
\textit{Cog-4:} Adjustable Speed & \cellcolor{colorR}R \tiny(UX) & \cellcolor{colorR}R & \cellcolor{colorC}C & \cellcolor{colorI}I & - &\cellcolor{phaseMidDev}Mid- Dev \\
\textit{Cog-5:} Hide Distractions & \cellcolor{colorR}R \tiny(UX) & \cellcolor{colorA}A & \cellcolor{colorC}C & \cellcolor{colorI}I & - &\cellcolor{phasePostLaunch}Post-Launch \\
\midrule
\textit{SH-1:} Visualize Sound & \cellcolor{colorR}R \tiny(UX) & \cellcolor{colorR}R & \cellcolor{colorC}C & \cellcolor{colorC}C & \cellcolor{colorA}A & \cellcolor{phaseEarlyDev}Early Dev \\
\textit{SH-2:} Multimodal Chat & \cellcolor{colorR}R \tiny(UX) & \cellcolor{colorA}A & \cellcolor{colorC}C & \cellcolor{colorI}I & \cellcolor{colorC}C & \cellcolor{phaseEarlyDev}Early Dev \\
\textit{SH-3:} Supplement Audio & \cellcolor{colorR}R \tiny(UX) & \cellcolor{colorA}A & \cellcolor{colorC}C & \cellcolor{colorC}C & \cellcolor{colorA}A & \cellcolor{phaseEarlyDev}Early Dev \\
\textit{SH-4:} Separate Volumes & \cellcolor{colorR}R \tiny(Audio/UX) & \cellcolor{colorR}R & \cellcolor{colorC}C & \cellcolor{colorC}C & \cellcolor{colorA}A & \cellcolor{phaseMidDev}Mid- Dev \\
\textit{SH-5:} Subtitle Settings & \cellcolor{colorR}R \tiny(UX) & \cellcolor{colorA}A & \cellcolor{colorC}C & \cellcolor{colorI}I & \cellcolor{colorC}C & \cellcolor{phasePostLaunch}Post-Launch \\ \bottomrule
\end{tabular}
\caption[]{\changed{Responsibility Assignment Matrix (RACI) with Implementation Phases. Roles are color-coded based on assignment: \colorbox{colorR}{\textbf{R}esponsible}, \colorbox{colorA}{\textbf{A}ccountable}, \colorbox{colorC}{\textbf{C}onsulted}, and \colorbox{colorI}{\textbf{I}nformed}. A dash (-) indicates no involvement necessary. Designer types shown in parentheses. The \textbf{Phase?} column shows when guidelines should be implemented: \colorbox{phaseDesign}{\textbf{Design}} (app architecture), \colorbox{phaseEarlyDev}{\textbf{Early Dev}} (prototyping/MVP), \colorbox{phaseMidDev}{\textbf{Mid- Dev}} (secondary features/polish), or \colorbox{phasePostLaunch}{\textbf{Post-Launch}} (maintenance). \\
\noindent\ \text{*}\textit{PM/Lead represents the equivalent role for a project manager at each organization's respective size: }i.e., \textit{Project/Product Manager (big tech/midsize), Technical Lead/Founder (startups), or self-coordinated/client-driven (freelancers).}}}
\Description{A responsibility assignment matrix (RACI) displaying 20 accessibility guidelines grouped into four categories: Motor (5 guidelines), Visual (5 guidelines), Cognitive (5 guidelines), and Sensory/Hearing (5 guidelines). Each guideline is assigned responsibility levels to five roles—Designers, Developers, QA/Test, PM/Lead, and Audio Engineer—using color-coded designations: Red (Responsible), Yellow (Accountable), Blue (Consulted), and Gray (Informed). A final column indicates the recommended implementation phase: Design, Early Dev, Mid-Dev, or Post-Launch. For example, "Mot-1: Remap Controls" assigns Designers as Accountable, Developers as Responsible, QA as Consulted, and PM as Informed, with implementation during the Design phase.}
\label{tab:raci}
\end{table*}}

\subsubsection{Implementation Timing and Responsibilities}
\label{a11ytiming} \label{sec:timing}
Practitioners' implementation preferences revealed clear patterns about when and who should integrate a11y into development workflows. \changed{We present these findings across all 20 guidelines using a responsibility assignment matrix (RACI)~\cite{AtlassianRACI} in Table~\ref{tab:raci} to help streamline complex projects with multiple stakeholders.}


\textbf{\textit{When should a11y be implemented?}} Participants overwhelmingly ($n=21$) felt a11y features should be considered and implemented at the design phase ($n=12$), project start ($n=4$), or early development ($n=3$) 
when possible, with safety-critical guidelines like \textit{Cog-1: Avoid Flicker} \textbf{requiring} careful planning due to inherent danger (P8). P8 believed it is worth a visual quality sacrifice to improve XR performance for this guideline: \textit{``It's literally everything. Because if you have a bad experience like this, number one, it could physically hurt somebody.''}
Other guidelines, like those for adapting user interfaces for people with visual impairments (\textit{Vis-2: Resizable UI}, \textit{Vis-4: Audio Description}, \textit{Vis-5: Text Voiceovers}), become ``prohibitively expensive'' late in development (P24): their structural, cross-system impact demands careful design decisions that would necessitate overhauling the system for a clean implementation. P6 emphasized this difficulty: \textit{``Trying to patch [Vis-5] on after the fact sounds like  disaster.''} 

\textbf{\textit{Who are responsible for implementing a11y guidelines?}} Participants mostly agreed ($n=18$) that \textit{designers} should be primarily responsible for implementing or enforcing a11y features, particularly guidelines affecting visual presentation and user experience (\textit{Vis-1: Supplement Color}, \textit{Vis-3: Readable Font Size}, \textit{Cog-2: Supplement Text}). In some cases, designers should be less involved in favor of developers (P19 for \textit{SH-2: Multimodal Chat}) or direct feedback from testers (P18 for \textit{SH-3: Supplement Audio}). Fourteen participants agreed that \textit{developers} should be primarily responsible for implementing more technical features, particularly system-level features like \textit{Mot-4: Simple Controls}, \textit{SH-4: Separate Volumes} , and \textit{Mot-2: Flexible Inputs}. 
We support prior work as practitioners taking on multiple roles often causes a11y to be deprioritized in smaller teams, potentially explaining why startup and midsize practitioners generally showed lower willingness to incorporate a11y features (P4, P7, P9, P20, P23).

\subsection{Community Visions for Guideline-Informed A11y Support} 
\label{findings-support} \label{sec:support}
Including and beyond evaluating guidelines, our findings reveal a critical need for developer-centered support tools that integrate with existing XR development workflows; specifically, a downloadable, free, open-source 3rd party a11y package or set of reusable components emerged as most preferred by practitioners ($n=21$), but further adoption barriers vary by organizational context.


\subsubsection{Third-Party Tool Adoption Patterns} \label{sec:tools}
Practitioners' willingness to adopt third-party a11y tools depends on organizational culture and technical constraints. Eleven participants noted using third-party packages in their XR projects, but adoption patterns differ even across these users.

Startup practitioners exhibit ``not-invented-here syndrome'' according to P2, with strict feelings against external packages and outside consultants perpetuating \textit{``This sort of philosophy that we know everything.''}
Other practitioners with big tech experience may be more likely to pick up 3rd party tools, quickly evaluating their ``worth'' towards speeding up their development or build time (P15, P19, P23). P23 emphasized a critical pitfall associated with third-party supports: \textit{``\textbf{You think the tool is giving you the time to implement it.} If a tool takes 25 hours to implement into your application, that's misery. If it takes 25 minutes, that's an actually good tool.''}
Big tech companies (P3, P15, P25) prefer internal tools or require open-source packages for security compliance. These companies also prefer code-based scene generation over visual editors to mitigate conflicts with version control; as P3 explained, \textit{``When someone changed the prefab, now my day's work is hosed. Because the stuff I did in the GUI with clicking buttons in 12 different places is gone.''}
This organizational variation suggests a11y tools must provide multiple integration paths: lightweight, open-source (and appropriately licensed) packages for larger companies, rapid prototyping tools for startups, and extensive customization options for specialized applications.


\subsubsection{Implementation Guidance and Examples} \label{sec:examples}
One of the greatest improvements to guidelines suggested by practitioners is to add specific a11y feature examples and their implementations ($n=20$) accompanying each guideline. 
These requests revealed three categories of desired support:

\textbf{Comparative examples:} Practitioners wanted both accessible and inaccessible implementations shown side-by-side in an ``easily digestible'' format (P12, P23), including videos or GIFs (P12, P19, P22) of PWD unable to use an app until the support method is added, potentially making the impact more evocative (P14). 

\textbf{Code artifacts:} Participants wanted downloadable project materials (P11) including Unity scenes (P3), code snippets with inline documentation (P8), and platform-specific implementations to be easily imported into preferred frameworks: OpenXR  ($n=7$), Unity XR Interaction Toolkit ($n=6$), and MRTK ($n=6$) users each requested examples in their platforms.

\textbf{Popular app references:} Five participants (P2, P4, P15, P21, P24) want guidelines to reference implementations in successful apps. We showcase the popular VR game Beat Saber's~\cite{beatsaber} use of colorblind options as an example in Appendix~\ref{appendix:beatsaber}.

\section{Discussion}



\begin{figure*}
    \centering
    \includegraphics[width=1\linewidth]{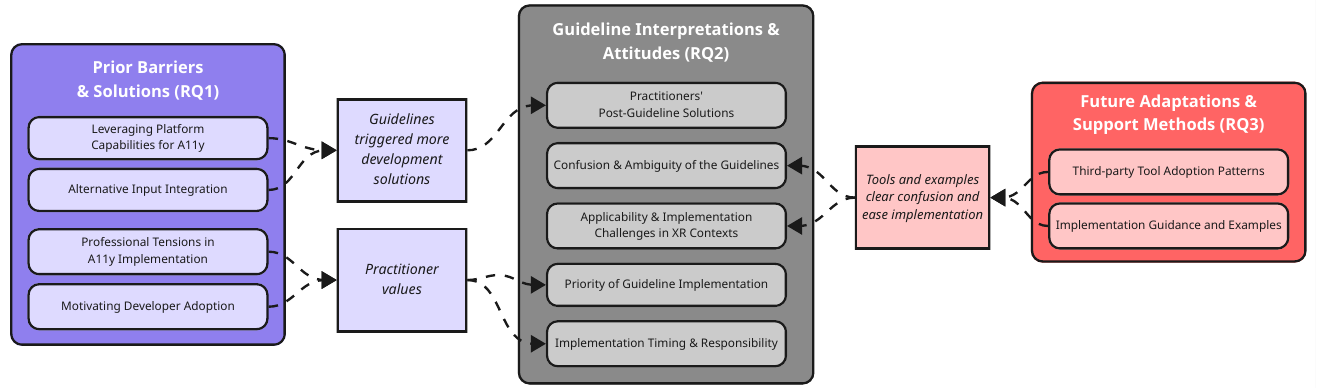}
    \caption[]{\changed{Mapping research questions addressed by our findings. RQ1 identified technical solutions, headset limitations, and training gaps. RQ2 revealed ambiguity in applying 3D guidelines to XR contexts and tensions between immersion and a11y. RQ3 showed practitioners need automated checking, concrete examples, and code-first support. Overall, practitioners are willing to implement a11y but need help balancing effort versus impact.}}
    \label{fig:researchquestions}
    \Description{Concept map showing three research questions and their findings. RQ1 (Prior Barriers and Solutions) identifies confusion in guidelines, implementation priority, platform capabilities, and alternative input integration. RQ2 (Guideline Interpretations and Attitudes) addresses professional tensions and practitioners' post-guideline solutions. RQ3 (Future Adaptations and Support Methods) shows implementation timing, XR-specific challenges, third-party tool adoption, and the need for implementation guidance and developer motivation. Arrows connect findings to show how they inform support methods and practitioner values.}
\end{figure*}


\changed{Our evaluation reveals that existing 3D a11y guidelines fail when applied to XR development realities; by examining how practitioners interpret, implement, and adapt existing guidelines, we establish that inadequate guidance, not just professional culture or organizational constraints, contributes to poor XR a11y adoption. Building on Wang et al.~\cite{wang2025understanding}'s work identifying general barriers to XR a11y implementation, our work provides the first practitioner-based assessment of specific guideline applicability to XR.

We document practitioners' existing a11y solutions and implementation barriers (RQ1), finding that practitioners develop creative workarounds independently, though these remain scattered without structured guidance. When presented with guidelines through structured discussions (RQ2), practitioners demonstrated strong technical ability by proposing concrete implementation strategies and identifying specific technical challenges, revealing both their capacity to translate abstract guidelines into practical approaches and identify the gaps that prevent successful application to XR contexts. However, this evaluation also revealed critical interpretation challenges: practitioners struggled to apply 3D virtual world guidelines to XR's unique spatial, temporal, and performance constraints, leading to ambiguity about scope and feasibility. Their willingness to adopt guidelines varied by organizational context, reflecting different perceptions of priorities and responsibilities. For support methods, practitioners strongly preferred platform-integrated tools and concrete code examples over static documentation (RQ3), revealing a mismatch between traditional a11y guidance formats and XR development workflows. These findings show that improving XR a11y requires rethinking how a11y knowledge is packaged and delivered to fit XR development practice. We reflect on these tensions in industry contexts in \S~\ref{disc:tensions} and derive actionable feedback for future support methods in \S~\ref{disc:informing}.}

\subsection{\changed{Professional Tensions Reflect Guidance Gaps}}
\label{disc:tensions}

Prior research has documented a11y challenges among software practitioners broadly~\cite{bi2022pract} and the gap between HCI research innovations and industry adoption~\cite{colusso2017gap}; our findings extend these insights to reveal how XR's unique demands create distinct tensions between immersive design values and a11y implementation. Our evaluation revealed that practitioners' struggles to balance immersion with a11y stem from inadequate guidance rather than inherent design conflicts (\S~\ref{sec:tensions-findings}). When practitioners express willingness to exclude users with disabilities to preserve their envisioned experience, or resist removing potentially harmful effects despite known safety risks, these decisions reflect a critical gap: existing guidelines fail to demonstrate how a11y and XR design can coexist.
Some current 3D guidelines recommend a11y features as additions or modifications to existing designs via separate modes, alternative interfaces, or removed features~\cite{apx,gameguide,xbox}. This framing positions a11y as potentially compromising the ``core'' experience, leaving practitioners to believe they must choose between creative vision and inclusive design, or dismiss a11y entirely as not ``worth'' their effort. However, this perceived trade-off appears to emerge from guidance limitations rather than technical necessity. 

\textbf{Guidelines Should Demonstrate Universal Benefits.} A11y guidelines should evolve from disability-specific checklists to frameworks demonstrating universal design benefits. As documented in \S~\ref{sec:motivating}, our participants recognized that features like alternate input methods and supplemental feedback improve experiences across user populations (even for users who may not identify as PWD, consistent with Microsoft's Inclusive Design Principles~\cite{microsoftInclusive101}). Guidelines should forefront these universal benefits while maintaining specific disability accommodations, helping practitioners understand a11y as \textit{expanding} rather than constraining design possibilities. This shift requires concrete examples of XR experiences achieving both immersion and broad a11y, but our participants' struggles to adapt these examples to diverse XR contexts reveal how a lack of guidance forces design choices that unnecessarily pit a11y against immersion.

\textbf{Platform-Integrated Support Tools.} Our practitioners strongly ($n=21$) preferred platform-integrated solutions over manual implementation (\S~\ref{sec:tools}). Therefore, XR development platforms should provide built-in a11y frameworks that maintain immersive quality. When platforms handle core a11y infrastructures like screen reader support, subtitle systems, and input remapping, developers can focus on design considerations while leveraging reliable a11y features. 
In cases where platforms don't implement a11y, recent \textit{post hoc} systems like Killough et al.'s \textit{VRSight}~\cite{killough2025vrsight} (that generates spatial audio scene descriptions for blind VR/MR users) may provide user-sided compensation without developer integration, demonstrating how technical workarounds can enable a11y without compromising immersive experiences.

\textbf{Education and Community Building.} Beyond guidelines and tools, addressing these tensions requires reframing a11y in practitioner culture. Our participants mostly relied on a11y implementations from popular titles (\S~\ref{sec:examples}), revealing how rarely accessible yet immersive XR experiences are publicly celebrated. The current guidance gap creates false choices between creative vision and inclusive design; filling this gap requires cultural transformation where a11y expertise becomes recognized as fundamental to XR design excellence rather than specialized knowledge only fulfilled for legal compliance.

\subsection{\changed{Informing Future XR A11y Guidelines}}
\label{disc:informing}
Research on translational resources demonstrates that practitioners prefer actionable, prescriptive guidance over abstract principles~\cite{colusso2017gap}, yet insufficient formal accessibility education often leaves practitioners ill-equipped to implement inclusive designs~\cite{putnam2023could}.
Building on our findings, we present design considerations for future XR a11y guidelines, addressing recurring patterns in how practitioners interpret, evaluate, and adopt a11y guidance 
within real-world XR development constraints. 
We frame these design insights as considerations rather than definitive recommendations, recognizing that guideline development involves complex trade-offs and that our practitioner sample, while diverse, may not capture all development contexts.

\textbf{\textit{1. Demonstrate Feasibility via Concrete Examples in XR Contexts}}. Practitioners familiar with web a11y encountered confusion applying concepts to XR, highlighting the need for concrete, XR-\textit{native} examples alongside abstract principles. 
\begin{itemize}
    \item[\textit{1.1}] \textit{Provide specific implementation values with citations} (\S~\ref{sec:confusion}). 
    Specify concrete values (\textit{e.g.,} safe flicker rates in Hz, minimum text sizes, readable distances) with authoritative citations.
    \item[\textit{1.2}] \textit{Include side-by-side visuals showing accessible vs. inaccessible implementations}  (\S~\ref{sec:examples}). Provide comparative examples through GIFs, videos, or screenshots demonstrating compliant and non-compliant designs.
    \item[\textit{1.3}] \textit{Reference successful implementations from popular apps} (\S~\ref{sec:motivating}). Cite examples from widely-recognized apps (\textit{e.g.,} Beat Saber's colorblind options, Appendix~\ref{appendix:beatsaber}) demonstrating feasibility and commercial viability.
    \item[\textit{1.4}] \textit{Provide downloadable sample scenes demonstrating first-person implementation}  (\S~\ref{sec:tools}). Offer complete examples as Unity and/or Unreal sample projects showing a11y features from the user's perspective (\textit{i.e.,} using an egocentric camera system like Unity's \textit{XR Rig}~\cite{xritool}).
\end{itemize}

\textbf{\textit{2. Establish Implementation Timing}}. A11y implementation difficulty and cost increase exponentially as projects mature, with post hoc additions often requiring fundamental architectural changes:
\begin{itemize}
    \item[\textit{2.1}] \textit{Implement a11y as early as possible during design phase} (\S~\ref{sec:timing}). Specify that a11y should be considered during initial project planning and architecture, emphasizing early integration reduces costs while acknowledging post hoc additions remain possible.
    \item[\textit{2.2}] \textit{Build a11y into experience design baselines before asset creation} (\S~\ref{sec:timing}). Establish a11y systems and frameworks before creating assets to ensure a11y shapes infrastructure rather than being constrained by existing designs.
\end{itemize}

\textbf{\textit{3. Address XR-Specific Constraints}}. XR's unique technical and interaction constraints (spatial tracking, kinesthetic interactions, headset limitations) require explicit guideline treatment:
\begin{itemize}
    \item[\textit{3.1}] \textit{Never attach UI elements to the camera or head} (\S~\ref{sec:applicability}). Head-locked UI causes motion sickness; specify world-anchored or body-anchored placement strategies.
    \item[\textit{3.2}] \textit{Account for headset fit variability affecting FOV, legibility, and comfort}  (\S~\ref{sec:applicability}). Specify adaptive placement strategies and generous safe zones for critical information across diverse users and headset configurations.
    \item[\textit{3.3}] \textit{Support locomotion via multiple methods}  (\S~\ref{sec:applicability}). Provide diverse locomotion options including physical movement, controller input, and alternate input methods appropriate for different XR application types.
    \item[\textit{3.4}] \textit{Distinguish AR vs. VR interaction requirements} (\S~\ref{sec:applicability}). Articulate AR vs. VR a11y requirement differences, including AR-specific occlusion handling and outdoor visibility.
    \item[\textit{3.5}] \textit{Test across multiple headset types to account for performance differences} (\S~\ref{sec:applicability}). Specify testing requirements across standalone and tethered headsets to address performance variations and platform-specific capabilities.
\end{itemize}

\textbf{\textit{4. Provide Implementation Tools}}. Practitioners require concrete tools and infrastructure beyond documentation to efficiently implement a11y features:
\begin{itemize}
    \item[\textit{4.1}] \textit{Offer free, open-source plugins and packages for major engines} (\S~\ref{sec:tools}). Provide pre-built, production-ready components for Unity and Unreal under open-access licenses to reduce implementation barriers.
    \item[\textit{4.2}] \textit{Include automated a11y checking tools and testing frameworks} (\S~\ref{sec:tools}). Develop automated a11y checking tools that identify common issues during development.
    \item[\textit{4.3}] \textit{Enable importing platform-level user settings} (\S~\ref{nativedevicehooks}). Provide mechanisms for XR apps to access users' OS a11y preferences, ensuring consistent experiences across apps.
    \item[\textit{4.4}] \textit{Provide a11y metadata tagging systems} (\S~\ref{sec:applicability}). Develop XR equivalents of alt text to enable developers to semantically describe 3D objects for assistive technologies.
    \item[\textit{4.5}] \textit{Follow OpenXR standards for cross-platform consistency} (\S~\ref{sec:tools}). Align a11y features with OpenXR specifications to ensure implementations work across diverse hardware platforms.
\end{itemize}

\textbf{\textit{5. Frame Motivation and Benefits}}. Practitioners need compelling justification for a11y investment given competing priorities and resource constraints:
\begin{itemize}
    \item[\textit{5.1}] \textit{Frame guidelines as general best XR design practices benefiting all users} (\S~\ref{sec:motivating}). Position a11y features as improvements to overall user experience rather than accommodations for a minority population.
    \item[\textit{5.2}] \textit{Quantify expanded market reach through user impact numbers} (\S~\ref{sec:motivating}). Provide concrete statistics on potential users affected by each a11y feature to justify development investment.
    \item[\textit{5.3}] \textit{Note that people develop disabilities over time} (\S~\ref{sec:motivating}). Emphasize that a11y benefits extend to aging, temporary impairments, and situational limitations.
    \item[\textit{5.4}] \textit{Highlight AAA\footnote{We use the term ``AAA'' in this instance to refer both to the games industry classification for high-budget titles produced by major publishers~\cite{bernevega2022assetization} 
    and to WCAG Level AAA, the highest conformance tier~\cite{w3web}.} a11y features as industry-viable} (\S~\ref{sec:examples}). Reference a11y in commercially successful titles demonstrating compatibility with high-quality, profitable XR experiences.
\end{itemize}

These considerations emerged from practitioner evaluation of existing guidelines and represent patterns we observed across diverse organizational contexts. However, we acknowledge these considerations require validation through guideline development and deployment studies. Different development contexts may require different approaches, and trade-offs between comprehensiveness and simplicity warrant further investigation.

\subsection{Limitations and Future Work}

Our study focused on evaluating existing guidelines rather than developing new XR-specific guidance. Without systematic practitioner evaluation of current resources, developing new standards would lack empirical grounding about real-world developer needs and constraints. Our evaluation strongly suggests that inadequate guidance, not just professional culture, contributes to systematic exclusion of PWD from XR experiences.

Given that this study is an a11y-focused work, we acknowledge potential recruitment bias toward a11y-supportive practitioners: many participants expressed interest in a11y and many had prior a11y experience; 
however, even if practitioners are particularly \textit{motivated} to implement a11y, we document significant technical barriers, guideline interpretation challenges, and implementation conflicts that prevent adoption. To enrich our data, future work should expand participant recruitment and consider lightweight methodologies (\textit{e.g.,} questionnaires) to attract broader developers in contributing their experiences. 

Our evaluation methodology provides a framework for assessing guideline effectiveness with practitioners, but requires application across broader practitioner populations and diverse XR application contexts. Future research should build on our foundational work to develop, test, and refine new XR-\textit{native} a11y guidelines that address the technical incompatibilities, organizational contexts, and implementation preferences we identified. We call for community contributors to add a11y samples and modules to open-source frameworks like \textit{XR Blocks}~\cite{Li2025XR}, contribute to open-source a11y tools like \textit{VRSight}~\cite{killough2025vrsight} and \textit{SeeingVR}~\cite{seeingVRtoolkitgithub}, and share their best practices and implementations built on existing game engines.


\section{Conclusion}


Through evaluating existing 3D a11y guidelines with 25 XR practitioners, we demonstrate that current guidelines systemically fail to address XR's unique technical requirements and development contexts. 
Practitioners prefer platform-integrated tools over written guidelines, requesting Unity-compatible packages, OpenXR integration, and early design-phase implementation. Our documented interpretation struggles and implementation barriers establish that inadequate guidance, not just professional culture, contributes to poor XR accessibility adoption.
\changed{Our findings reveal that guidelines addressing timing controls appeared technically feasible but fundamentally inappropriate for real-time XR experiences, while platform-specific challenges like standalone headset performance constraints and spatial audio complexity remained entirely unaddressed by current 3D virtual world guidelines. For guideline creators, our work emphasizes the need for concrete implementation examples with XR-specific technical specifications, clear responsibility assignments across organizational contexts, and frameworks that address XR's unique constraints--- spatial interaction paradigms, real-time performance requirements, and novel input methods--- rather than simple adaptations from 2D or desktop 3D contexts.}
Our evaluation provides essential evidence for developing XR a11y support that addresses real practitioner needs while enabling the inclusive immersive experiences that XR technology promises. 

\begin{acks}
This work was partially supported by the National Science Foundation under Grant No. IIS-2328182. Thank you to everyone who has helped us in publishing this manuscript over the years, including but not limited to: Our participants for their valuable insight; to Justin Feng and Daniel Wang for their assistance aggregating interview data; and to our labmates for their feedback. 
\textbf{For an extended \\version of this work, please see Killough et al. ``XR for All''~\cite{killough2024xrall}.}
\end{acks}

\bibliographystyle{ACM-Reference-Format}





\appendix

\begin{table*}[t]
\section{Example Codes}
\label{appendix:themes}
  \centering
  \hyphenpenalty=1000
\fontsize{7.3}{9.3}\selectfont
  \begin{tabular}{>{\raggedright\arraybackslash}p{0.17\textwidth}>{\raggedright\arraybackslash}p{0.24\textwidth}>{\raggedright\arraybackslash}p{0.54\textwidth}}
  \toprule
  \textbf{Themes} & \textbf{Sub-Themes} & \textbf{Example Codes} \\
  \midrule
  \textbf{Difficulties of Feature Integration} & A11y Issues and Needs in XR Interactions & Text illegible due to resolution; need ARIA label equivalent for XR; spatial audio for BLV; thumbstick turning harms mental map; VR motion sickness; vision-motion mismatch; no linear page flow in 3D; real-time caption challenges; lacking VR screenreader \\
  \cmidrule{2-3}
  & Sacrificing A11y for Immersion & Balance a11y and immersion; ignore guidelines for creative direction; a11y solutions could make experience worse \\
  \cmidrule{2-3}
  & Disability Representation in XR & Normative view; perception: PVI can't use VR; perception: apps don't accommodate PWD \\
  \cmidrule{2-3}
  & Performance Limitations of Standalone HMDs & Performance overhead to add a11y feature; hardware limitations; additional restrictions for mobile VR; Android vs PC environment differences \\
  \cmidrule{2-3}
  & Onboarding First-Time Users & Tag objects as interactable; video tutorial insufficient; trouble onboarding to XR; VR inexperience \\
  \midrule
  \textbf{Current Development Practices} & Lack of Formal XR Training & Self-taught; community resources; holes in development knowledge; StackOverflow; Reddit; Discord; YouTube; Udemy; Udacity; blog posts \\
  \cmidrule{2-3}
  & Developers' Diverse Responsibilities & Developers wear many hats; survival mode; high turnover \\
  \cmidrule{2-3}
  & Dynamic Code Generation & Code easier to merge vs prefabs; dynamically create objects by script; procedurally generate data \\
  \cmidrule{2-3}
  & Reusing Code Between Projects & Reusing code; reuse components; base template scene; investing in reusable a11y frameworks \\
  \cmidrule{2-3}
  & Testing with PWD & PWD appear in general testing; difficult to find PVI testers; no access to testers;  dedicated user studies \\
  \midrule
  \textbf{Motivations} & Motivations & Time; money; audience size worth; legal; government regulations; good PR; fear of bad reviews; a11y helps people without disabilities \\
  \cmidrule{2-3}
  & Hindrances & Survival mode; MVP; a11y low priority; a11y cut to meet deadlines; conflicting priorities; limited resources; ``not realistic'' to add a11y \\
  \cmidrule{2-3}
  & Shared Responsibility & Accessible design; best design practices; including accessible design lowers costs; everyone has a responsibility to make accessible; no dedicated a11y developer; a11y should not be a hard sell to VR devs \\
  \midrule
  \textbf{Practitioners' Current XR A11y Solutions}   & Examples of Existing A11y Solutions & Colorblindness filters; one-handed mode; text size; UI magnification; variable height experience; visual sound indicator; volume sliders; invisible buttons \\ 
\cmidrule{2-3}
    & Incorporating Alternate Input & Rewired; Xbox a11y controller; Leap Motion; Joycon; Kinect; Merge Cube; XRInput \\
  \cmidrule{2-3}
  & Hooking into System A11y Support & Custom hooks; Unity lacks a11y hooks; hook into platform functionality; screenreader; TTS; STT; invisible buttons \\
  \midrule
  \textbf{XR-Specific Applicability Concerns} & Prior Knowledge of Guidelines & Current guidelines broad; WCAG; Microsoft's MR Standards; lack of XR guidelines; lack of awareness \\
  \cmidrule{2-3}
  & Understanding Guidelines & Not for real-time gaming; 2D not applicable to 3D; trying to fit old a11y requirements into XR; semantically ambiguous \\
  \cmidrule{2-3}
  & When to Implement a11y & A11y in design phase; a11y should start asap; easier the earlier you start; difficult to incorporate post-hoc \\
  \midrule
  \textbf{Preferred XR A11y Support Methods} & SDK's/Toolkits & Unity XR Interaction Toolkit; MRTK; OpenXR support; Meta/Oculus toolkit \\
  \cmidrule{2-3}
  & Guideline Shortcomings & Semantically ambiguous; undefined evaluation metrics; guideline needs technical scope; WCAG is 2D-only; no guideline consensus; Microsoft's MR Accessibility Standards not fleshed out \\
  \cmidrule{2-3}
  & 3rd Party Packages & Skeptical of 3rd party packages; criteria: actively developed, lightweight, open-source, easy to drop into projects, customizable \\
  \cmidrule{2-3}
  & Game Modification & Modding security  concerns; modding legal problems; big tech dislikes/disallows modding; offload responsibility to community; multiplayer concerns \\
  \cmidrule{2-3}
  & Automated A11y Checking & Lacking automated a11y checks for 3D; AI a11y checking; use 3rd party a11y checker \\
  \cmidrule{2-3}
  & Guideline Improvement Suggestions & Provide examples; specific values; cost/time estimates; sample scenes; GIFs/Videos; safe vs dangerous examples; implementation demonstrations \\
  \cmidrule{2-3}
  & Need to be Directly Implementable & Easy to implement; drag and droppable plugin; plug and play; limited overhead to add a11y \\
  \bottomrule
  \end{tabular}
\caption{Non-Exhaustive Summary of Themes, Sub-themes, and Corresponding Example Codes}
\Description{A three-column table categorizing qualitative coding findings with six main themes related to XR accessibility development. The themes are: Difficulties of Feature Integration (covering accessibility issues in XR interactions, balancing accessibility with immersion, disability representation, performance limitations, and user onboarding); Current Development Practices (including lack of formal training, diverse developer responsibilities, code generation, code reuse, and testing with people with disabilities); Motivations (covering motivating factors, hindrances, and shared responsibility for accessibility); Practitioners' Current XR accessibility Solutions (describing existing accessibility solutions, alternate input methods, and system accessibility support); XR-Specific Applicability Concerns (addressing guideline knowledge, understanding, and implementation timing); and Preferred XR accessibility Support Methods (covering SDKs/toolkits, guideline shortcomings, third-party packages, game modification, automated checking, and improvement suggestions). Each theme contains multiple sub-themes with corresponding example codes extracted from interview data.}
\label{table:themes}
\end{table*}

\clearpage

\onecolumn

\section{Beat Saber Color Options}
\label{appendix:beatsaber}

\begin{figure*}[h]
    \centering
    \includegraphics[width=0.95\linewidth]{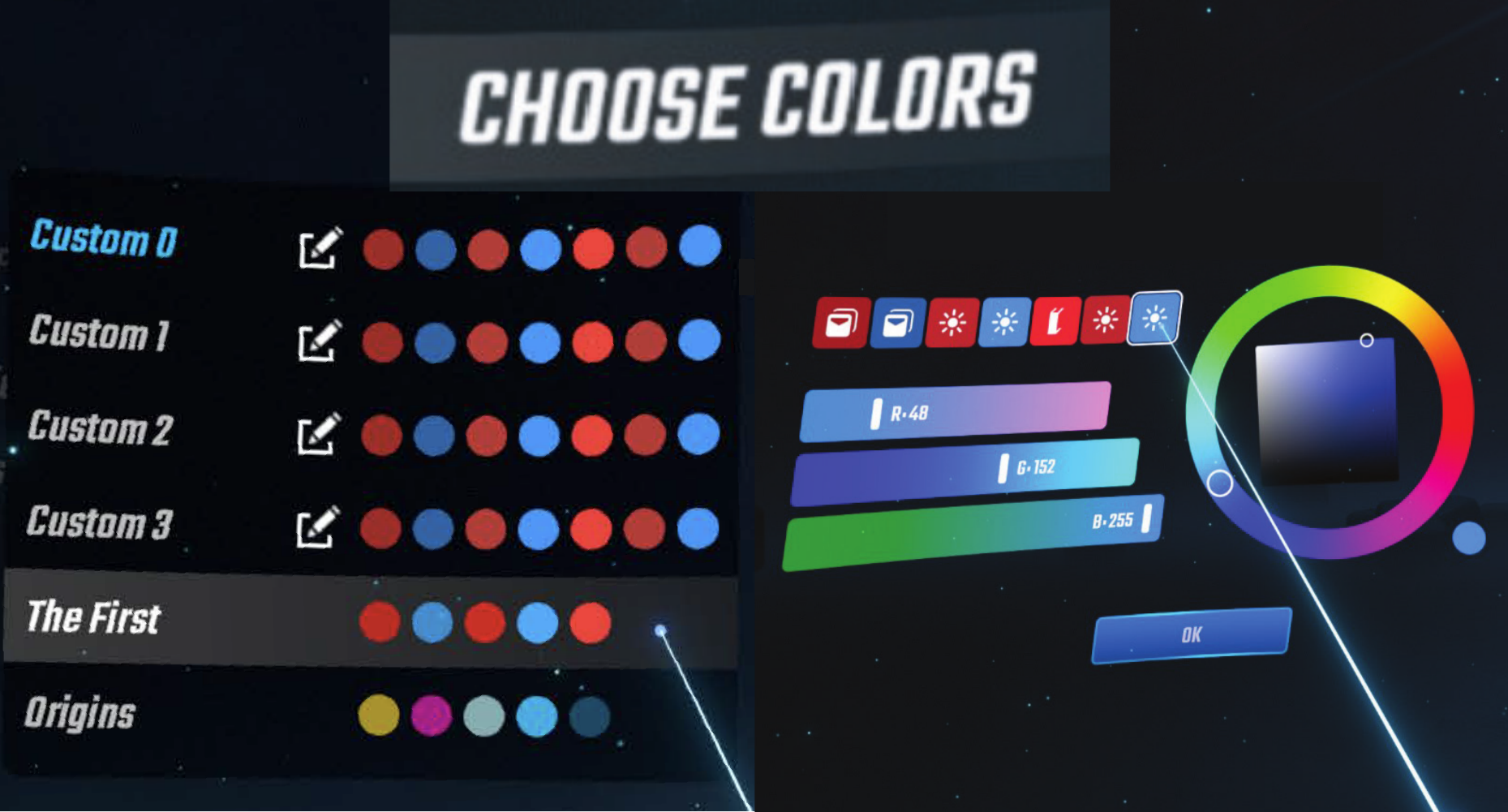}
    \caption[]{We show an example from one of the most popular~\cite{faric2019players} VR games, Beat Saber~\cite{beatsaber}, which offers custom color settings for a variety of virtual objects. Selecting an object (Right) like notes, lights, or wall, then picking a color from the picker, will cause the respective virtual object to change to that color. Users can create up to four custom color palettes (Left, labeled Custom 0 through Custom 3) or choose from a variety of presets themed from in-game collections (e.g., ``The First'', ``Origins''). As of November 2024 there do not exist built-in presets for common color vision deficiencies (e.g., protanopia, deuteranopia, tritanopia)~\cite{colorblindnesstypes}.}
    \Description{Edited screenshots from the VR game Beat Saber displaying the in-game color picker. The top of the image reads "Choose Colors". The left half of the image shows six color palette options. The first four read "Custom zero" through "Custom three", and the last two read "The First" and "Origins". All six options are selectable. The four "custom" options are editable. Each color palette is shown with five to seven circles next to it, where every one except "Origins" shows a Red and Blue color scheme. Origins shows an orange, pink, and blue color scheme. On the right half of the image is a series of buttons next to a color picker, with a blue OK button at the bottom. Seven buttons on top show icons with two notes, two lights, one wall, then two more notes. each set of two alternates red then blue. Selecting one icon changes the color picker to that color. The color picker is a red, green, and blue slider on the left side, and a circular wheel on the right side. A square sits in the middle of the wheel to pick saturation that is colored in a gradient, with dark blue on the right side and gray on the left, white in the top left corner and black on the bottom. A virtual reality pointer points to the color palette option called "The First" on the left half of the figure, and another virtual reality pointer points to the rightmost blue light icon.}
    \label{fig:supports}
\end{figure*}

\end{document}